%% file: _gl411.tex
\documentclass[twocolumn]{aastex63}
\usepackage[T1]{fontenc}
\usepackage{amsmath}
\usepackage{tabularx, booktabs}
\usepackage{graphicx}
\usepackage{natbib}
\usepackage[scaled]{helvet}
\usepackage{epsfig}
\usepackage{url}
\bibpunct{(}{)}{;}{a}{}{,}
\usepackage{textcomp}
\usepackage{xspace}
\usepackage{apjfonts}
\usepackage{rotating}
\usepackage{gensymb}
\usepackage{numprint}
\npthousandsep{,}
\usepackage{color}
\usepackage[normalem]{ulem}
\usepackage[left]{lineno}

\input{newcommands.tex}

\input{variables.tex}

\input{affiliations.tex}

\begin{document}

\title{Confirmation of the Long-period Planet Orbiting Gliese 411 and the Detection of a New Planet Candidate}

\input{authors.tex}

\shorttitle{Gliese 411 Planetary System}
\shortauthors{Hurt et al.}

\begin{abstract}

We perform a detailed characterization of the planetary system orbiting the bright, nearby M dwarf Gliese 411 using radial velocities gathered by APF, HIRES, SOPHIE, and CARMENES. We confirm the presence of a signal with a period near 2900 days that has been disputed as either a planet or a long-period stellar magnetic cycle. An analysis of activity metrics including the $\mathrm{H_\alpha}$ and $\mathrm{log'R_{HK}}$ indices supports the interpretation that the signal corresponds to a Neptune-like planet, GJ 411 c. An additional signal near 215 days was previously dismissed as an instrumental systematic, but we find that a planetary origin cannot be ruled out. With a semi-major axis of $\adavgerr$ au, this candidate's orbit falls between those of its companions and is located beyond the outer edge of the system's habitable zone.\footnote{Determined using the moist greenhouse and maximum greenhouse limits in \cite{Kopparapu:2013}} It has a minimum mass of $\msinidavgerr$ \mearth, giving a radial-velocity amplitude of $\kdavgerr$ \ms. If confirmed, this would be one of the lowest-amplitude planet detections from any of these four instruments. Our analysis of the joint radial-velocity data set also provides tighter constraints on the orbital parameters for the previously known planets. Photometric data from \tess\ do not show any signs of a transit event. However, the outermost planet and candidate are prime targets for future direct imaging missions, and GJ 411 c may be detectable via astrometry.

\vspace{.8cm}
\end{abstract}

\section{Introduction}

The discovery of planets orbiting nearby stars has traditionally relied on the radial-velocity (RV) detection method. Out of the 89 known planets within 10 pc of the solar system, 86 were discovered---and each has been observed---using this method. In contrast, less than a fifth of all known planets were discovered using Doppler spectroscopy.\footnote{The data used in this paper can be accessed via the NASA Exoplanet Archive link \dataset[10.26133/NEA1]{https://exoplanetarchive.ipac.caltech.edu/cgi-bin/TblView/nph-tblView?app=ExoTbls&config=planets}.}

Many of these radial-velocity detections were made through systematic monitoring such as the Eta-Earth Survey \citep{Howard:2009}, which used the HIRES spectrograph at the W.M. Keck Observatory \citep{Vogt1994} to find low-mass ($\mysim3-30$ \mearth) planets around 230 of the closest GKM stars. The California Planet Search (CPS) group has continued similar observing campaigns, including the APF-50 Survey \citep{Fulton:2016, Fulton:2017:PhD}, which uses the robotic Automatic Planet Finder (APF; \citealt{Radovan:2014, Vogt:2014}) to observe the 51 brightest and least chromospherically active stars in the Eta-Earth Survey at high cadence. Most of these stars have been observed over baselines spanning more than one or two decades, offering the opportunity to detect long-period planets. Moreover, high-resolution spectroscopy of nearby stars can be obtained with shorter exposures, allowing many precise RV measurements that make it easier to detect low-amplitude signals. As a consequence, these surveys have played an important role in understanding the frequency of low-mass planets, along with their characteristics \citep{Howard:2010b, Mayor:2011, Fulton:2017:PhD}.

Gliese 411, also known as Lalande 21185 or HD 95735, is one of the brightest and closest M dwarfs to our solar system \citep{Lapine:2011, Gaidos:2014}, making it a frequent target in the search for exoplanets. Early astrometric studies suggested the presence of an orbital companion with estimates of a minimum mass ranging from $0.1$ to $0.3$ \msun\ \citep{vandeKamp:1951, Lippincott:1960}. Also using astrometry, \cite{Gatewood:1996} later claimed that a planet with a minimum mass of $0.9$ \mjup\ was orbiting the star. None of these claims have since been verified. Over the past five years, several RV searches for planets around GJ 411 have been published, beginning with \cite{Butler:2017}, who used HIRES spectroscopy to identify a $9.9$ day signal corresponding to a minimum mass of $3.8$ \mearth. This signal, however, has been disputed by subsequent analyses \citep{Diaz:2019}. Using RVs from the SOPHIE spectrograph \citep{Perruchot:2008}, \cite{Diaz:2019} discovered GJ 411 b, a planet with a minimum mass of $3$ \mearth\ on a $12.95$ day orbit. \cite{Stock:2020} subsequently combined the SOPHIE data with RVs from the CARMENES spectrograph \citep{Quirrenbach:2014, Quirrenbach:2018} to confirm this planet. They also identified a $2900$ day periodicity, which they attributed to long-period magnetic activity cycles. The California Legacy Survey also detected GJ 411 b and the 2900 day signal but instead attributed the latter to a planet with a minimum mass of $18$ \mearth, GJ 411 c \citep{Rosenthal:2021}. 

In this paper, we combine additional RVs gathered by the Eta-Earth and APF-50 Surveys with data from \cite{Diaz:2019}, \cite{Stock:2020}, and \cite{Rosenthal:2021} to conduct a detailed search for planets around GJ 411. We find that the disputed long-period signal is inconsistent with the star's magnetic activity cycle and favor the planetary interpretation. We investigate another signal identified by \cite{Rosenthal:2021} as an instrumental systematic and show that a planetary origin cannot be ruled out. Additionally, we provide new constraints on the orbital parameters of GJ 411 b and the most detailed characterization of the system to date. In \rfsecl{observations}, we describe our observations and RVs. We provide an updated analysis of the star's properties in \rfsecl{stellar}. In \rfsecl{data}, we describe our analysis of the RVs. We then discuss the transit, direct imaging, and astrometric detection prospects in \rfsecl{discussion} and place the system in the broader exoplanet population. Lastly, in \rfsecl{conclusions}, we conclude our results.

\section{Observations and Archival Data}
\label{sec:observations}

\subsection{HIRES Spectroscopy}
\label{sec:hires}

We obtained high-resolution spectra of GJ 411 with the HIRES spectrograph located on the Keck I telescope \citep{Vogt1994} at Maunakea, HA. The star was observed with a resolving power of $R\approx\numprint{60000}$ over wavelengths ranging roughly from $300$ nm to $1000$ nm. Spectra were wavelength calibrated using a warm iodine-gas cell and then modeled using a deconvolved stellar spectral template, an atlas iodine spectrum, and an approximate line-spread function. We derived RVs following the techniques described by \cite{Butler:1996}. We additionally calculated the $\mathrm{logR'_{HK}}$ and $\mathrm{H_\alpha}$ indices for each spectra, following the procedures outlined in \cite{Isaacson:2010} and \cite{Robertson:2014}.

While our observations originally ranged from June 1997 to January 2021, HIRES received a new CCD and other optical improvements in 2004. Before these upgrades, charge transfer efficiency in the CCD would blur the dense spectral lines found in the spectra of M dwarfs like GJ 411, contaminating the RVs. Consequently, we only include the HIRES data gathered after these upgrades, leaving 299 spectra beginning in November 2004. Five observations had low counts, resulting in decreased signal to noise and inflated uncertainties; we discarded these measurements, giving a total of 294 radial velocities. Our HIRES observations up to 2014 are shared with \cite{Butler:2017} but processed through a separate pipeline. Additionally, all RVs collected prior to April 2020 are shared with \cite{Rosenthal:2021}.

\subsection{APF Spectroscopy}
\label{sec:apf}

We collected 435 spectra of GJ 411 using the Automated Planet Finder (APF; \citealt{Radovan:2014, Vogt:2014}) between March 2014 and February 2021. The APF is a robotic telescope located at Lick Observatory on Mt. Hamilton, CA. It is equipped with the Levy Spectrograph, a high-resolution echelle spectrometer that achieves $R\approx\numprint{100000}$ over wavelengths ranging from $374.3$ nm to $980$ nm. Spectra passed through a warm iodine-gas cell for wavelength calibration and were modeled the same way as the HIRES spectra in \rfsecl{hires}. Radial velocities were then derived with the pipeline described by \cite{Fulton:2015}, which originates from the \cite{Butler:1996} pipeline used for the HIRES observations. We also calculated the $\mathrm{logR'_{HK}}$ and $\mathrm{H_\alpha}$ indices for each APF spectrum. Thirteen of the observations had low counts and were excluded from our analysis, leaving 422 RVs. Similar to the HIRES observations, all APF data from before April 2020 were included in \cite{Rosenthal:2021}'s analysis. The new APF and HIRES RVs are provided along with their uncertainties and respective activity indices in \rftabl{rvs}.

\input{rvs}

\subsection{SOPHIE Spectroscopy}
\label{sec:sophie}

\cite{Diaz:2019} collected 155 spectra of GJ 411 between October 2011 and June 2018 using the SOPHIE spectrograph \citep{Perruchot:2008, Perruchot:2011} located on the 1.93 m reflector telescope at the Haute-Provence Observatory in France. The observations were made in high-resolution mode ($R\approx\numprint{75000}$) and span wavelengths ranging from $387.2$ nm to $694.3$ nm. In addition to radial velocities, they derive observables such as $\mathrm{logR'_{HK}}$ and $\mathrm{H_\alpha}$ indices for each spectra, although they note the latter are influenced by an unknown systematic present in those data for other intensely observed stars. We obtained these measurements using the Vizier catalog access tool. The original RVs provided had not been corrected for instrumental zero-point changes; however, the appropriate offsets were calculated from a set of constant-RV stars \citep{Courcol:2015, Hobson:2018} and had been provided along with the rest of the publicly available data. After applying these offsets, there still appeared to be a discrepancy between the RVs and the data presented in \cite{Diaz:2019}; we concluded that they were not corrected for secular acceleration. We accounted for this discrepancy using the model described by \cite{Kurster:2003}, which relies on the star's measured distance, proper motion, and absolute radial velocity. Informative priors on the distance and proper motion were placed using data from Gaia Early Data Release 3 (see \rftabl{sparams}; \citealt{Gaia:2016B, Gaia:2020b}). After fitting this model, the residuals are used throughout the remainder of this work as the SOPHIE RVs. Two RVs had uncertainties greater than three times the median uncertainty and were removed, leaving 153 data points.

\input{stellarparams}

\subsection{CARMENES Spectroscopy}
\label{carmenes}

\cite{Stock:2020} collected 321 spectra of GJ 411 using the CARMENES spectrograph \citep{Quirrenbach:2014, Quirrenbach:2018}, installed at the 3.5 m telescope of the Calar Alto Observatory in Spain. Radial velocities were derived from the VIS channel, which has a resolution of $R\approx\numprint{95000}$ and a wavelength coverage of $520$ nm to $960$ nm. We accessed the publicly available RVs through Vizier, along with the respective $\mathrm{H_\alpha}$ indices, although we note these are calculated using a different method from the APF and HIRES indices \citep{Schoefer:2019}. The differential line width (dLW) and chromatic index (CRX) for each observation were also provided. Several RVs had large error bars and we removed any where the uncertainties were greater than three times the median uncertainty, leaving 309 data points. 

\subsection{TESS Photometry}
\label{tess}

GJ 411 was observed by NASA's \tess\ mission \citep{Ricker:2015} during sector 22, between February 19 and March 17 of 2020. We downloaded the short-cadence data file from the Mikulski Archive for Space Telescopes (MAST) and analyzed the presearch data conditions simple aperture photometry (PDCSAP) light curve, which has been corrected for instrumental systematics \citep{stumpe:2012, Smith:2012, Stumpe:2014}\footnote{The TESS data presented in this paper can be accessed via the MAST link \dataset[10.17909/t9-ejma-t702]{https://doi.org/10.17909/t9-ejma-t702}.}.

\section{Stellar Properties}
\label{sec:stellar}

Gliese 411 is an M1.9 star \citep{Mann:2015} located in the constellation of Ursa Major. The star is one of our closest neighbors at a distance of $2.5451\pm0.0002$ pc \citep{Gaia:2016B, Gaia:2020b}. It also has a V-band magnitude of $7.49$, making it the third-brightest red dwarf \citep{Lapine:2011, Gaidos:2014}.

Following the procedures outlined in \cite{Petigura:2017} and \cite{Dedrick:2020}, which utilize the publicly available packages {\tt SpecMatch-Empirical} \citep{Yee:2017} and {\tt isoclassify} \citep{Huber:2017, Berger:2020}, we derive the effective temperature, mass, radius, age, metallicity, and log-gravity for the star from our HIRES and APF spectra. Using the available values from each instrument, we also find that the star has a median $\mathrm{logR'_{HK}}$ of $-5.47\pm0.10$, suggesting that it experiences very low activity levels. However, previous photometric observations have shown rotational modulation with a period of $56.15\pm0.27$ days \citep{Diaz:2019}. Each of these stellar properties can be found in \rftabl{sparams}.

\section{Analysis of Radial-velocity Signals}
\label{sec:data}

\subsection{Signal Discovery}
\label{sec:discovery}

\begin{figure}
    \centering
    \includegraphics[width=\linewidth]{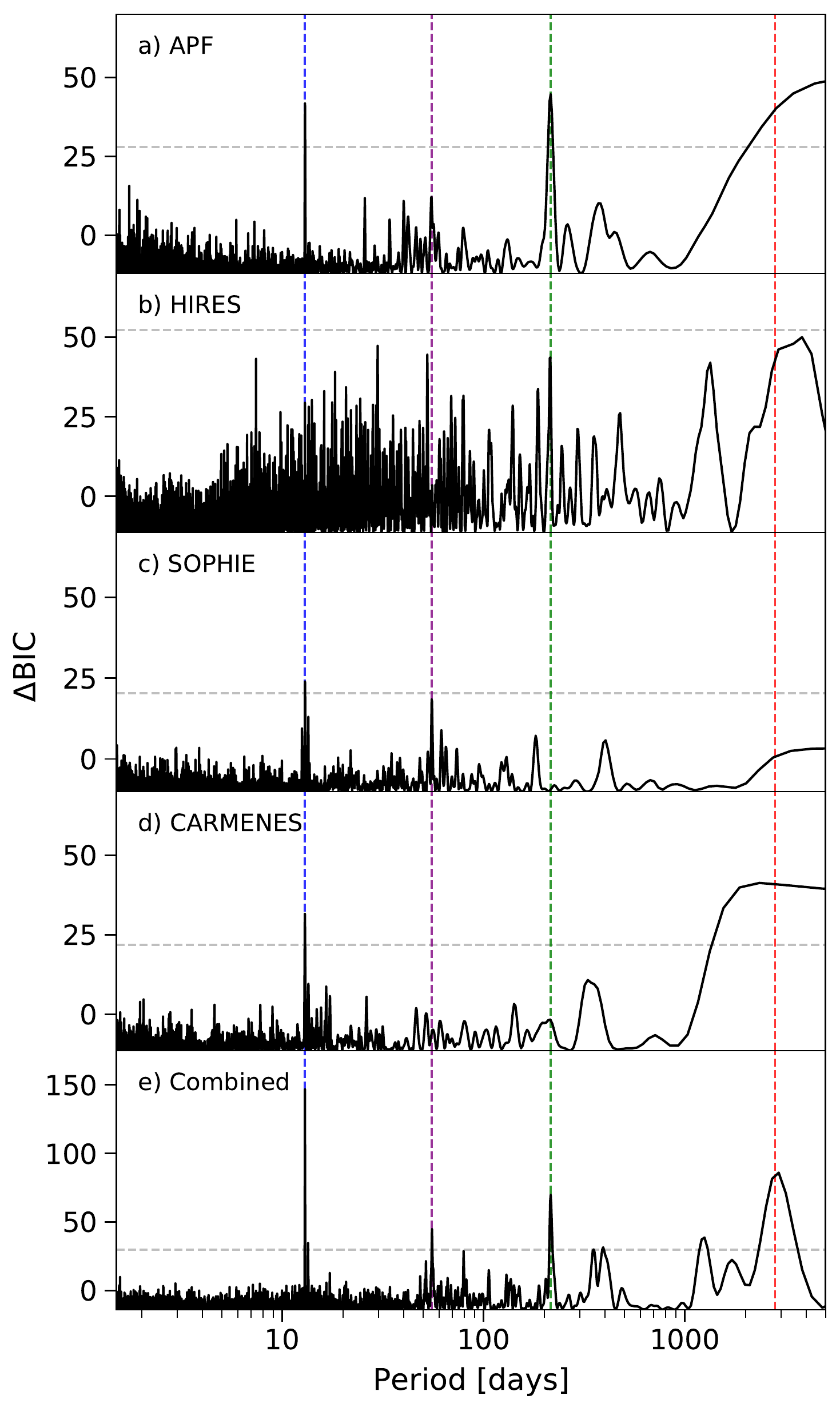}
    \caption{$\Delta$BIC periodograms for each data set (first four panels) and the combined radial velocities (bottom panel). The horizontal dashed grey lines mark the $0.1\%$ FAP threshold for each periodogram. The vertical dashed lines mark the periods of interest, with blue corresponding to GJ 411 b ($\mysim12.9$ days), red denoting the long-period, disputed signal ($\mysim2900$ days), green marking the candidate signal ($\mysim215.2$ days), and purple indicating the rotation period ($\mysim 55.3$ days).}
    \label{fig:periodograms}
\end{figure}

We generate $\Delta$BIC periodograms using {\tt RVSearch} \citep{Rosenthal:2021} to identify periodic signals in each data set along with the combined RVs, shown in \rffigl{periodograms}. This type of search is preferred to generalized Lomb-Scargle periodograms (GLS; \citealt{Zechmeister:2009}) because it accounts for instrument-specific parameters such as jitter and offset, allowing us to more accurately search the data from all four instruments simultaneously. Several peaks with false-alarm probabilities (FAPs) under $0.1\%$ appear in the periodogram of the combined data, with the most prominent located at $12.9$ days and corresponding to the known planet GJ 411 b. This signal is easily identified separately in the APF, SOPHIE, and CARMENES data but surprisingly is not visible in the HIRES periodogram. It is possible that HIRES, the lowest-resolution instrument out of all four, blurred the dense spectral lines of the M dwarf, contributing to increased noise and uncertainty in the RVs. This, combined with the data set's relatively low cadence, could hide a low-amplitude (<2 \ms) signal observed by the other instruments such as GJ 411 b.

The second peak is located at 2923 days, corresponding to the disputed signal identified by \cite{Stock:2020} and \cite{Rosenthal:2021}. In \rfsecl{distinct}, we investigate whether the peak is caused by stellar magnetic cycles. The APF and CARMENES periodograms show significant power near this periodicity but plateau because their baselines are too short to constrain any long-period signals. While insufficient precision prevents HIRES from making a strong detection of the signal, a somewhat resolved peak appears in the corresponding periodogram. Two marginally significant peaks located near 340 days and 420 days appear to be one-year aliases of this long-period signal and disappear, along with a peak near 1300 days, when fitting a Keplerian to a period near 2900 days. This 1300 day peak is not well-distinguished from the long-period signal, and we conclude it is likely a harmonic or alias, given that our data set's baseline is unable to well-constrain the 2900 day periodicity. 

The third peak in the combined RVs is located at $215.2$ days and was also identified by \cite{Rosenthal:2021}, but was dismissed as an APF systematic, primarily because it was not detected in the HIRES data. We note that while not significant, there is a resolved peak exactly at this period in the HIRES periodogram, although the signal is not independently detected in the SOPHIE or CARMENES data either. We further investigate this periodicity's origins in \rfsecl{systematic} and will refer to it as the candidate signal.

A fourth peak is found at $55.3$ days, which we note is approximately one-fourth of $215$ days and could easily be mistaken for a harmonic of the candidate signal. However, considering that \cite{Diaz:2019} found the stellar rotation period to be $56.16$ days using well-sampled photometry and that this peak is surrounded by a forest of smaller peaks, this most likely corresponds to rotational modulation and is consistent with the known distribution of rotation periods for cool stars \citep{McQuillan:2014}. Out of the four data sets, this modulation is most significant in the SOPHIE RVs.

\subsection{Isolating the Magnetic Activity Cycle}
\label{sec:distinct}

\begin{figure*}
    \centering
    \includegraphics[width=.9\linewidth]{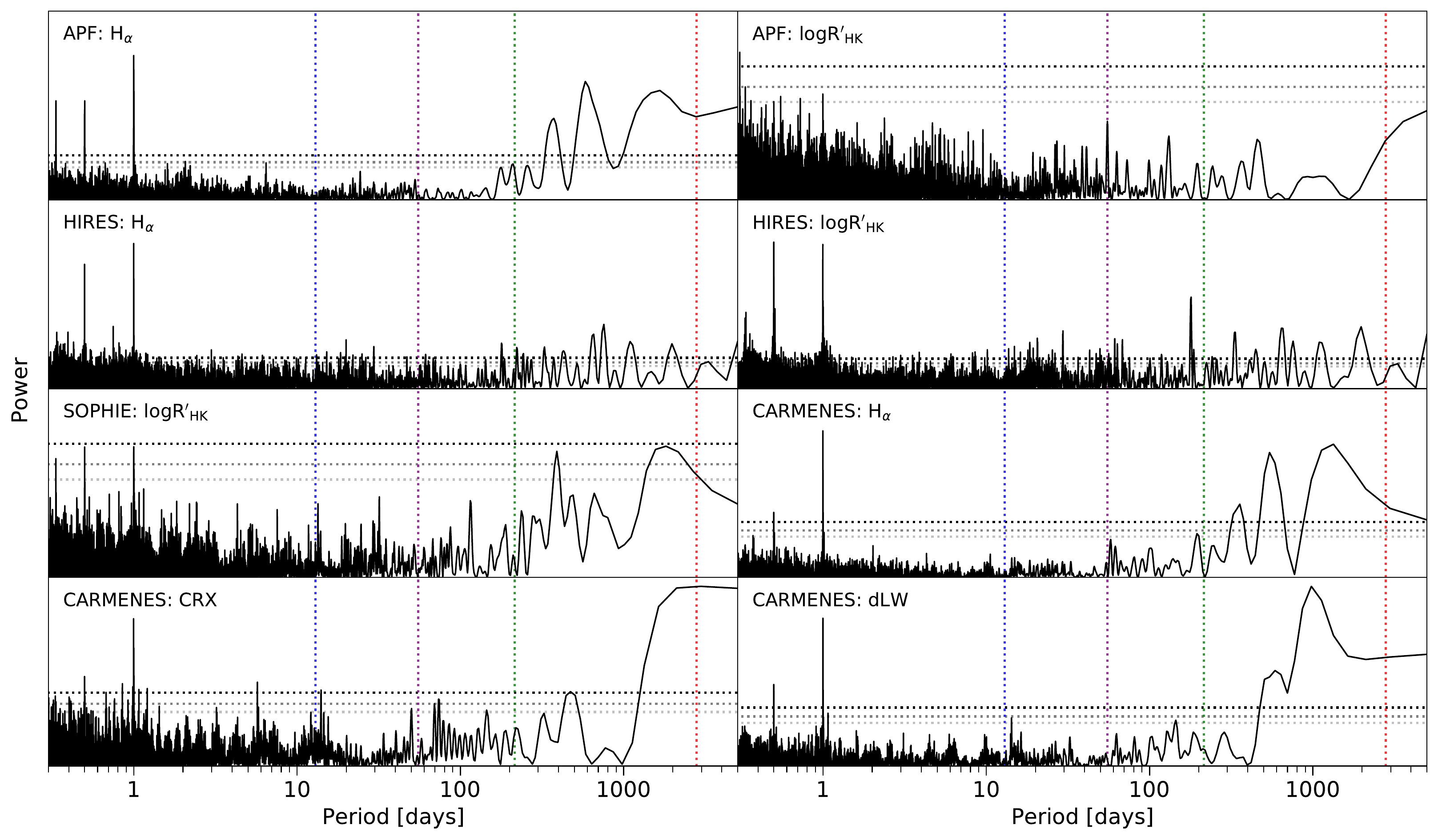}
    \caption{GLS periodograms of the activity indices for each instrument. The three horizontal dashed lines in each panel represent the $1\%$, $0.5\%$, and $0.01\%$ FAP levels for the corresponding periodogram. As with \rffigl{periodograms}, the vertical dashed lines represent the signals of interest, with blue corresponding to GJ 411 b, purple representing the rotational modulation, green marking the candidate signal, and red denoting the long-period disputed signal. No significant or resolved peaks appear near these four periodicities; however periodograms for the APF, SOPHIE, and CARMENES indices do show elevated power near the long-period signal of $\mysim2900$ days. This structure likely appears because the baselines for these data sets are too short to constrain any long periods.}
    \label{fig:actper}
    \vspace{.4cm}
    \includegraphics[width=.9\linewidth]{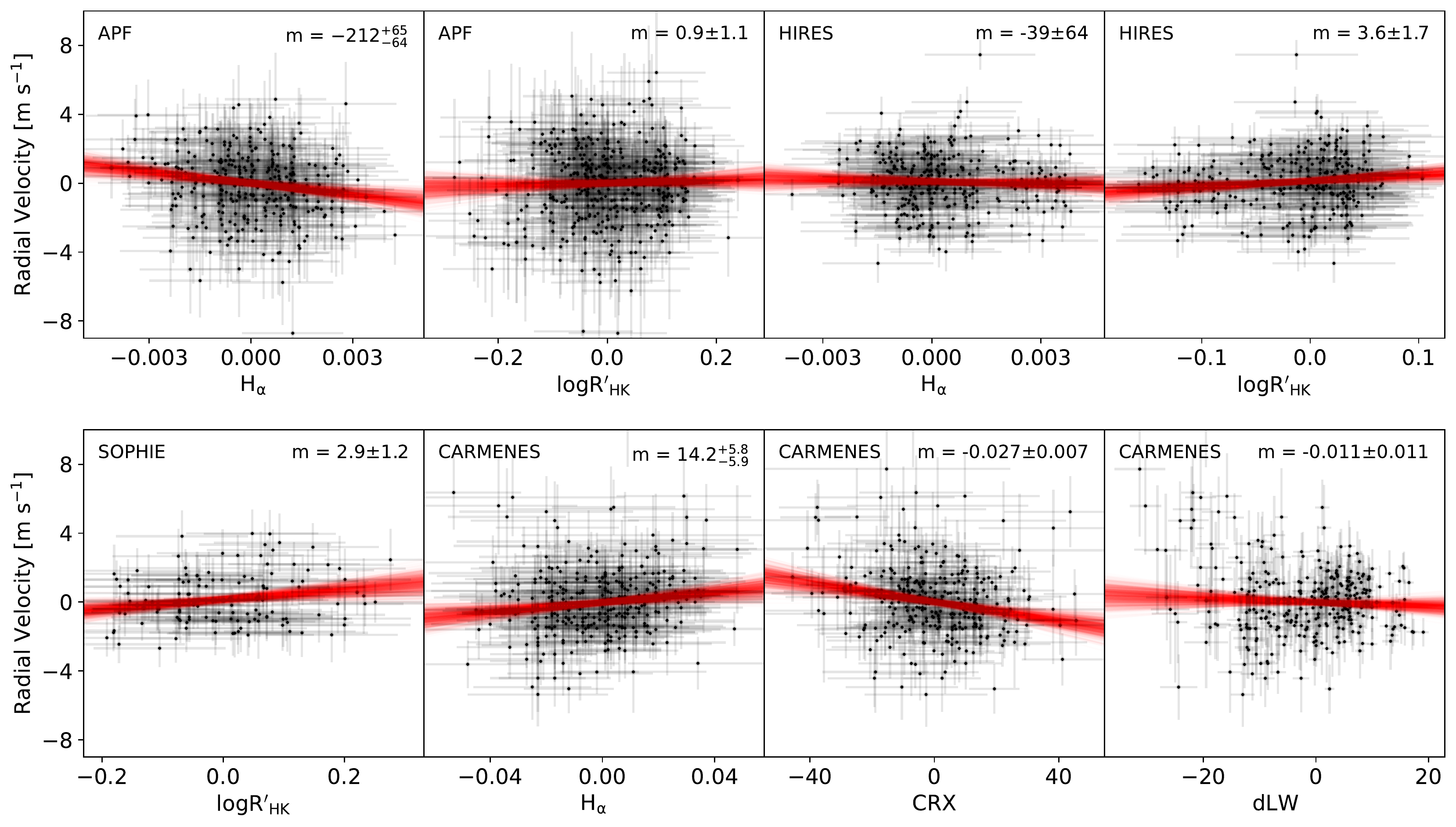}
    \caption{Plots of the radial velocities versus activity indices for each data set. Signals corresponding to GJ 411 b, the candidate, and stellar rotation have been subtracted from the RVs using the model presented in \rfsecl{model}. Samples of linear models from the MCMC posterior for each pair of variables are shown as red lines. The $68.3\%$ confidence interval for the slope is given in the top-right corner of each subplot. No correlations are observed for the APF and HIRES $\mathrm{logR'_{HK}}$ values while the rest of the correlations appear to be weak and can be explained by non-astrophysical signals or stellar magnetic cycles that span even longer than $2900$ days.}
    \label{fig:actcorr}
\end{figure*}

\begin{figure}[!ht]
    \centering
    \includegraphics[width=\linewidth]{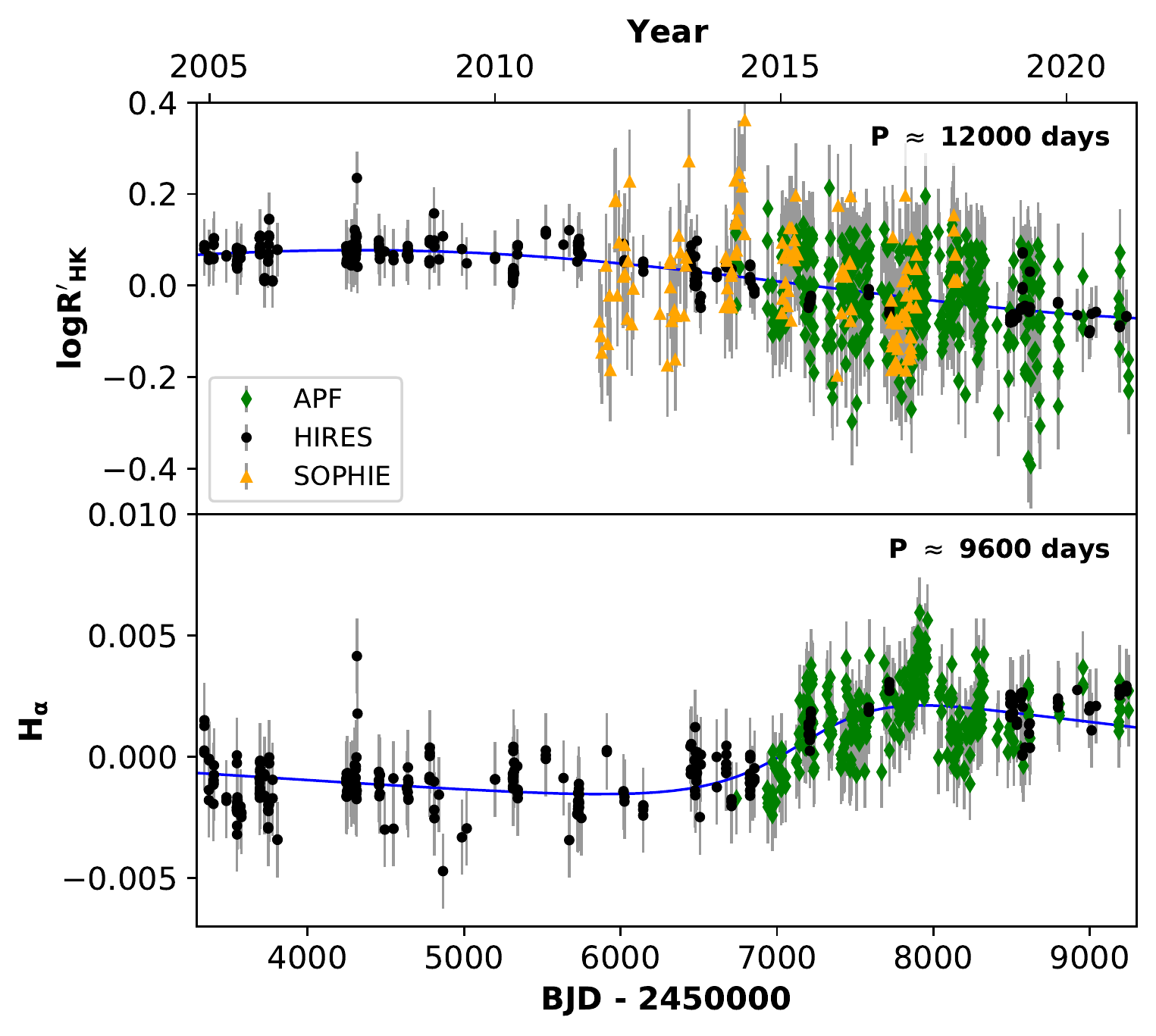}
    \caption{The time series data and best-fit Keplerian models for the $\mathrm{logR'_{HK}}$ (top) and $\mathrm{H_\alpha}$ (bottom) activity indices. We observe less than one cycle of the long-period signals contained in both, whereas our radial-velocity data span nearly two cycles of the $2900$ day signal (see \rffigl{rv} for comparison). This indicates that the disputed signal does not correspond to stellar magnetic activity and instead originates from a planet designated as GJ 411 c.}
    \label{fig:longterm}
\end{figure}

\cite{Stock:2020} first identified a long-period signal of $2852\pm568$ days using the SOPHIE and CARMENES data. They primarily use activity metrics derived from the CARMENES spectra to vet the origins of this periodicity, claiming that GLS periodograms show elevated power near 2900 days. We reproduce these periodograms for the CARMENES $\mathrm{H_\alpha}$, CRX, and dLW indices along with observables from the other instruments in \rffigl{actper}. While periodograms of the APF, SOPHIE, and CARMENES activity indices are all elevated near $2900$ days, they also plateau and do not resolve any peaks, indicating that they have baselines far too short for a periodogram analysis to constrain the origins of a long periodicity. \cite{Stock:2020} also identify several peaks near $1400$ days, approximately half of the signal in question. While our periodogram analysis also shows these peaks for the CARMENES $\mathrm{H_\alpha}$ and dLW metrics, this period is suspiciously close to the length of time spanned by the CARMENES data.

\cite{Rosenthal:2021} independently identified a signal of $3190\pm185$ days, consistent with the long periodicity from \cite{Stock:2020}, using their HIRES and APF RVs. They note that their $\mathrm{S_{HK}}$ indices for HIRES, which span a baseline longer than the disputed signal, show signs of a long-term trend. However, periodograms of the indices reveal no power near 2900 days. Similarly, our periodograms of the HIRES $\mathrm{H_\alpha}$ and $\mathrm{logR'_{HK}}$ indices do not contain significant power in this region. This indicates that any detectable magnetic activity cycle must be much longer and lends credence toward a planetary origin.

Because our periodogram analysis is limited by the baseline of each instrument, we also search for correlations between the radial velocities and stellar activity metrics for each data set, following the procedure outlined in \cite{Robertson:2014}. We are most interested in whether or not the long period is caused by activity, so we remove the signals corresponding to GJ 411 b, the candidate, and stellar rotation from our RVs using the model presented in \rfsecl{model}. We also reject any outlying activity indices with a three-sigma clip and then offset each data set such that their respective means are zero, mitigating degeneracies between slope and intercept when searching for linear trends. To determine whether the RVs and activity metrics are correlated, we fit a linear model accounting for uncertainties in both variables along with intrinsic scatter \citep{Hogg:2010} using a Markov chain Monte Carlo (MCMC) process. If a slope of zero is within one sigma of the resulting posterior, we consider there to be no correlation. Plots showing representative samples of the posteriors for each activity index can be found in \rffigl{actcorr}. Only two indices, the APF and SOPHIE $\mathrm{logR'_{HK}}$ values, do not show any correlation. However, we note that all of the correlations appear to be weak, with the actual data showing large amounts of scatter. Further, these relationships seem to be independent of the 2900 day signal. While periodograms of the HIRES $\mathrm{H_\alpha}$ and $\mathrm{logR'_{HK}}$ values show no power near 2900 days, both of the indices are correlated with the RVs, implying some other signal must be driving these relationships, such as stellar magnetic cycles even longer than $2900$ days. Although the CARMENES $\mathrm{H_\alpha}$ values are derived differently from the APF and HIRES $\mathrm{H_\alpha}$ values, both are related to the line width, and we expect them to behave similarly. We see that the APF and HIRES $\mathrm{H_\alpha}$ indices are negatively correlated with the RVs while the CARMENES $\mathrm{H_\alpha}$ indices are positively correlated, meaning some of the observed relationships could be non-astrophysical. While unable to provide a conclusive answer, our correlation analysis provides no strong evidence in favor of stellar activity as an origin of the disputed signal. 

Visually, there is a long-term trend present in the $\mathrm{logR'_{HK}}$ and $\mathrm{H_\alpha}$ indices. We use {\tt RadVel} \citep{fulton:2018a} to jointly model the APF, HIRES, and SOPHIE $\mathrm{logR'_{HK}}$ values with a single Keplerian. The best-fit model returns a period of $\mysim\numprint{12000}$ days; however, because this signal spans such a long time, MCMC fits are poorly constrained and fail to converge. Similarly, we fit a Keplerian model to the APF and HIRES $\mathrm{H_\alpha}$ indices, returning a best-fit period of $\mysim\numprint{9600}$ days. These models and the corresponding time series are shown in \rffigl{longterm}. Interestingly, the $\mathrm{H_\alpha}$ and $\mathrm{logR'_{HK}}$ indices appear to be anti-correlated, consistent with observations for other low-activity M dwarfs \citep{Gomes:2011}. It is apparent that any long-period magnetic activity cycle must be several times longer than the $2900$ day signal and could be driving the weak correlations observed between the activity metrics and RVs. Consequently, we agree with with \cite{Rosenthal:2021}'s interpretation that the disputed long-period signal originates from a planet, GJ 411 c.

\begin{figure*}[!t]
    \centering
    \includegraphics[width=0.48\linewidth]{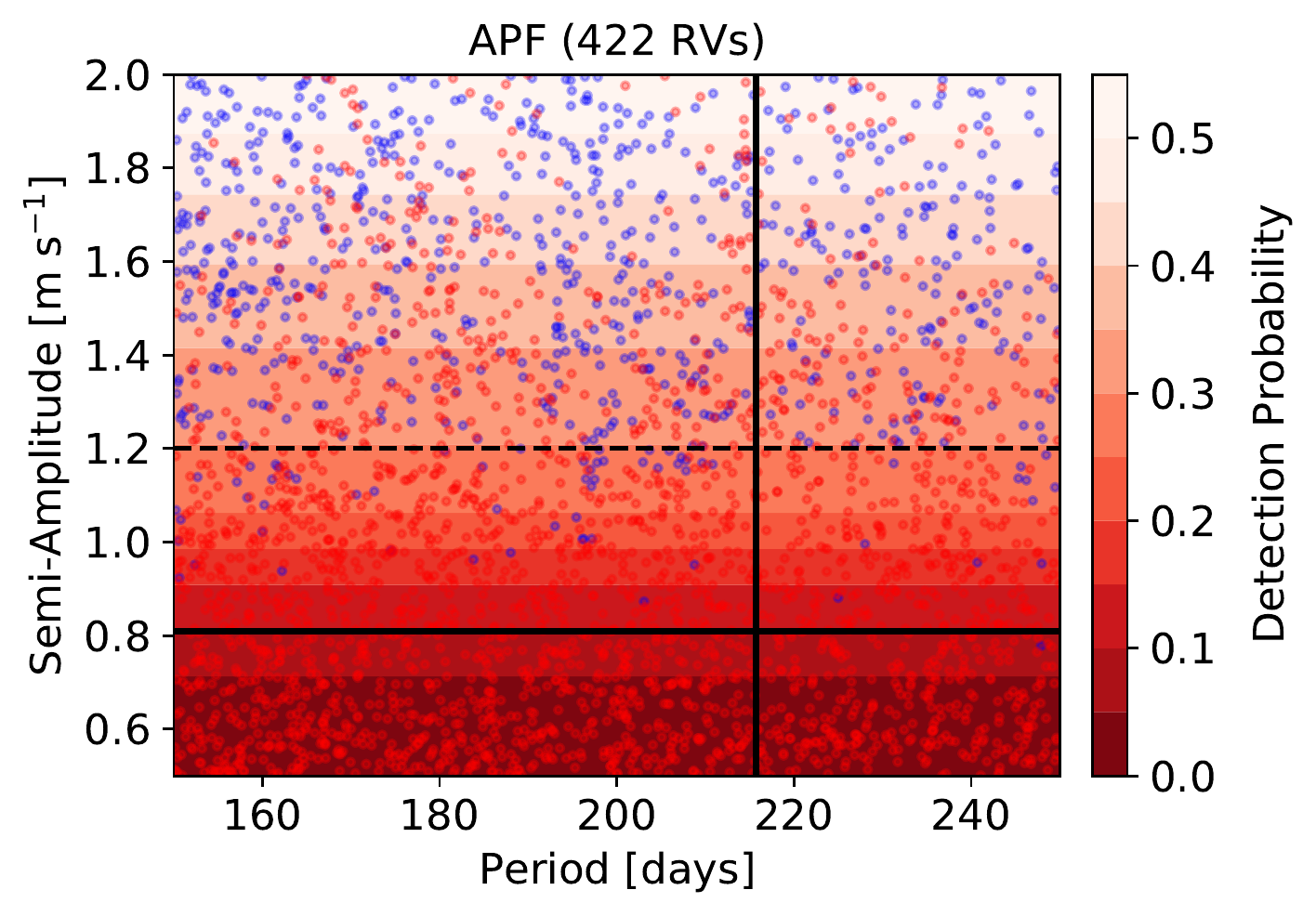} 
    \hspace{0.2cm}
    \includegraphics[width=0.48\linewidth]{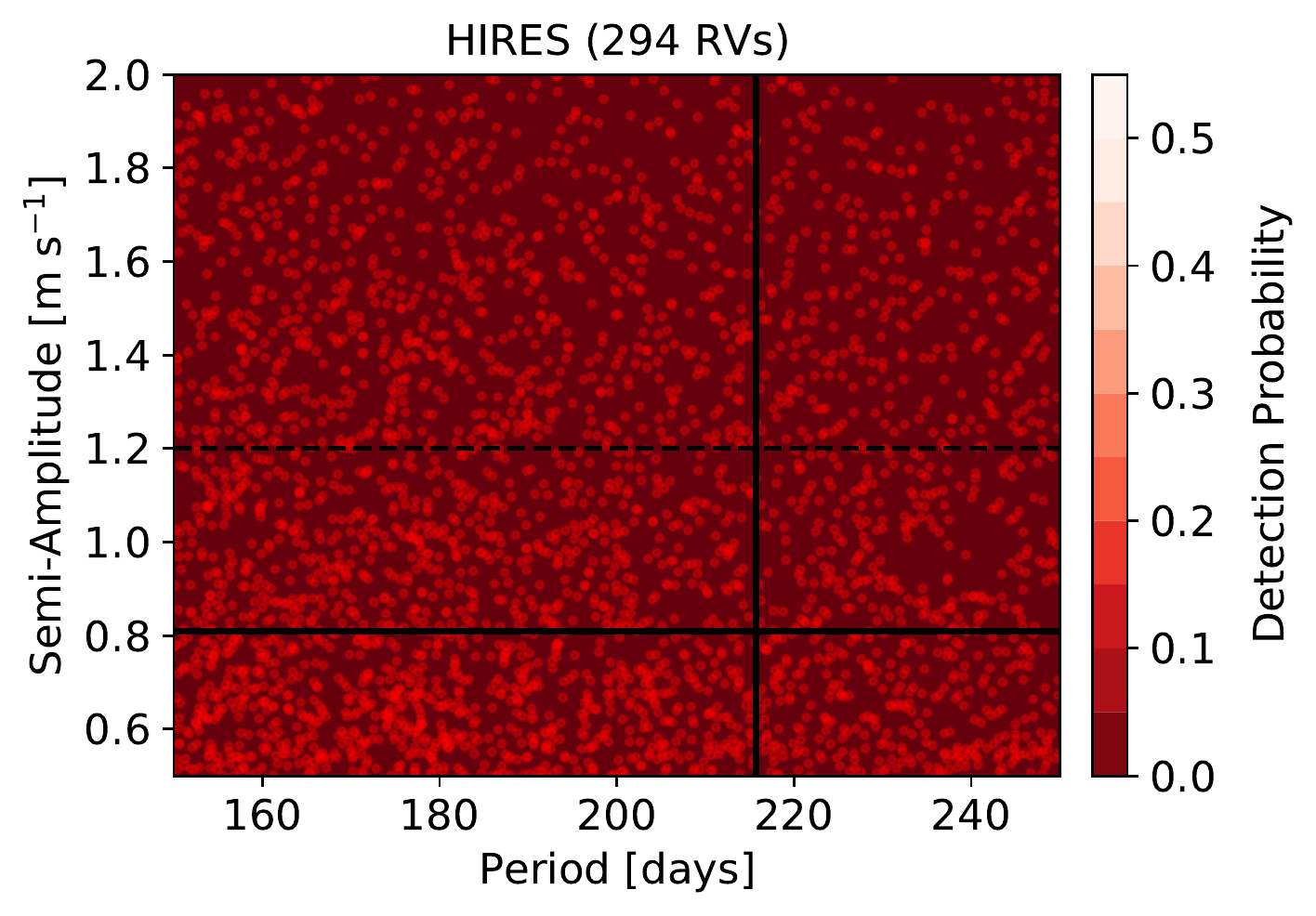}\\
    \vspace{0.07cm}
    \includegraphics[width=0.48\linewidth]{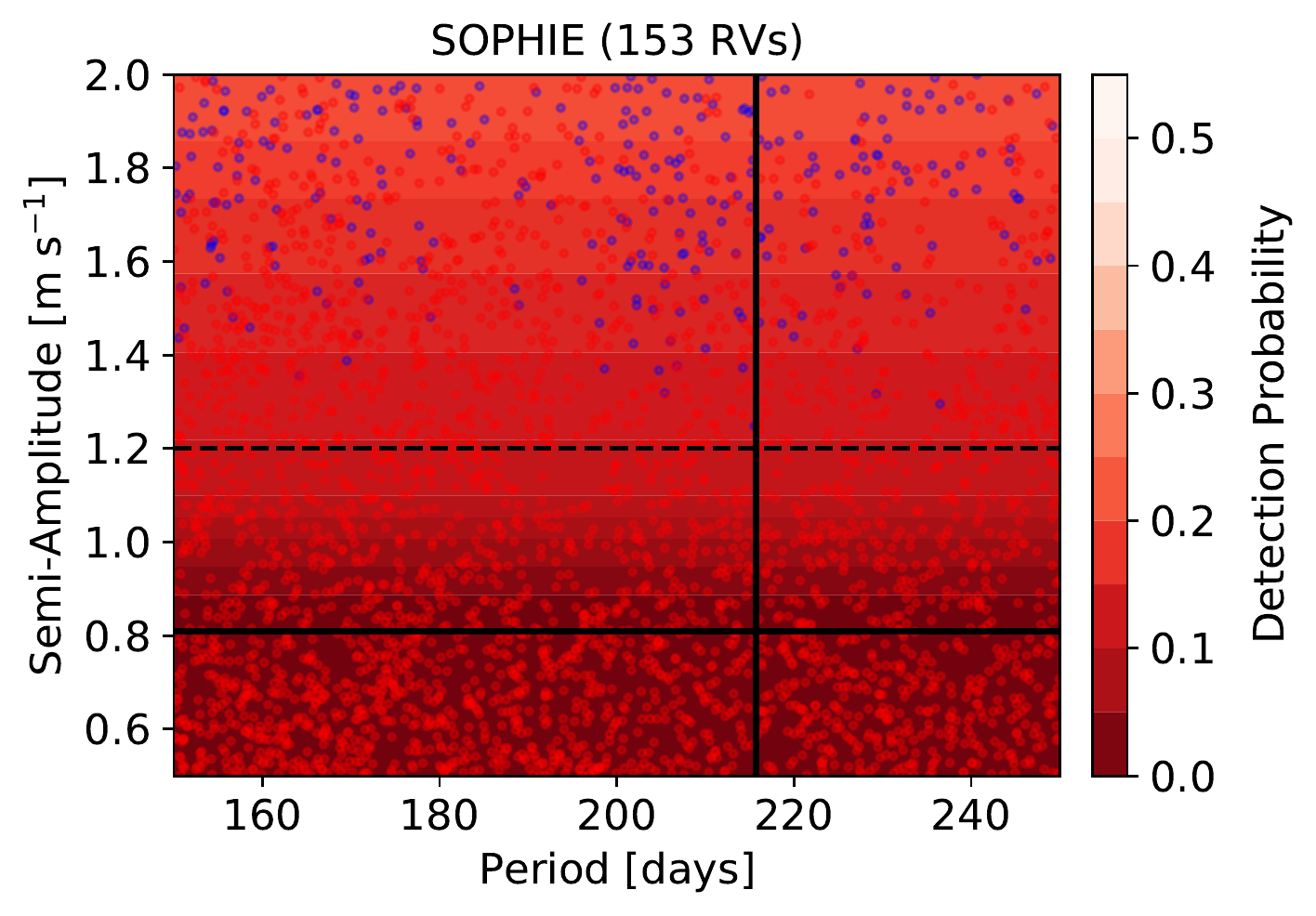}
    \hspace{0.2cm}
    \includegraphics[width=0.48\linewidth]{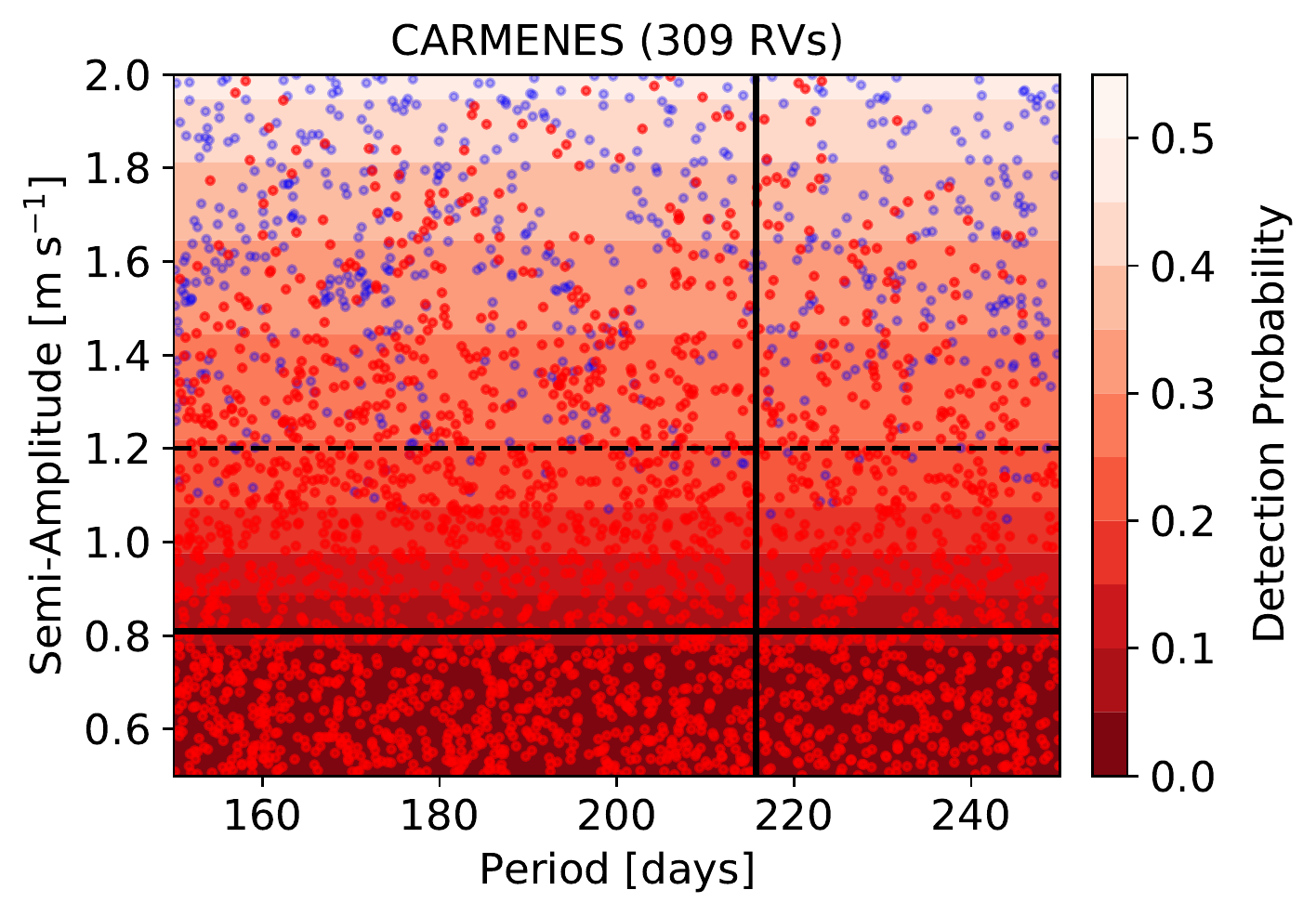}\\
    \vspace{0.07cm}
    \includegraphics[width=0.48\linewidth]{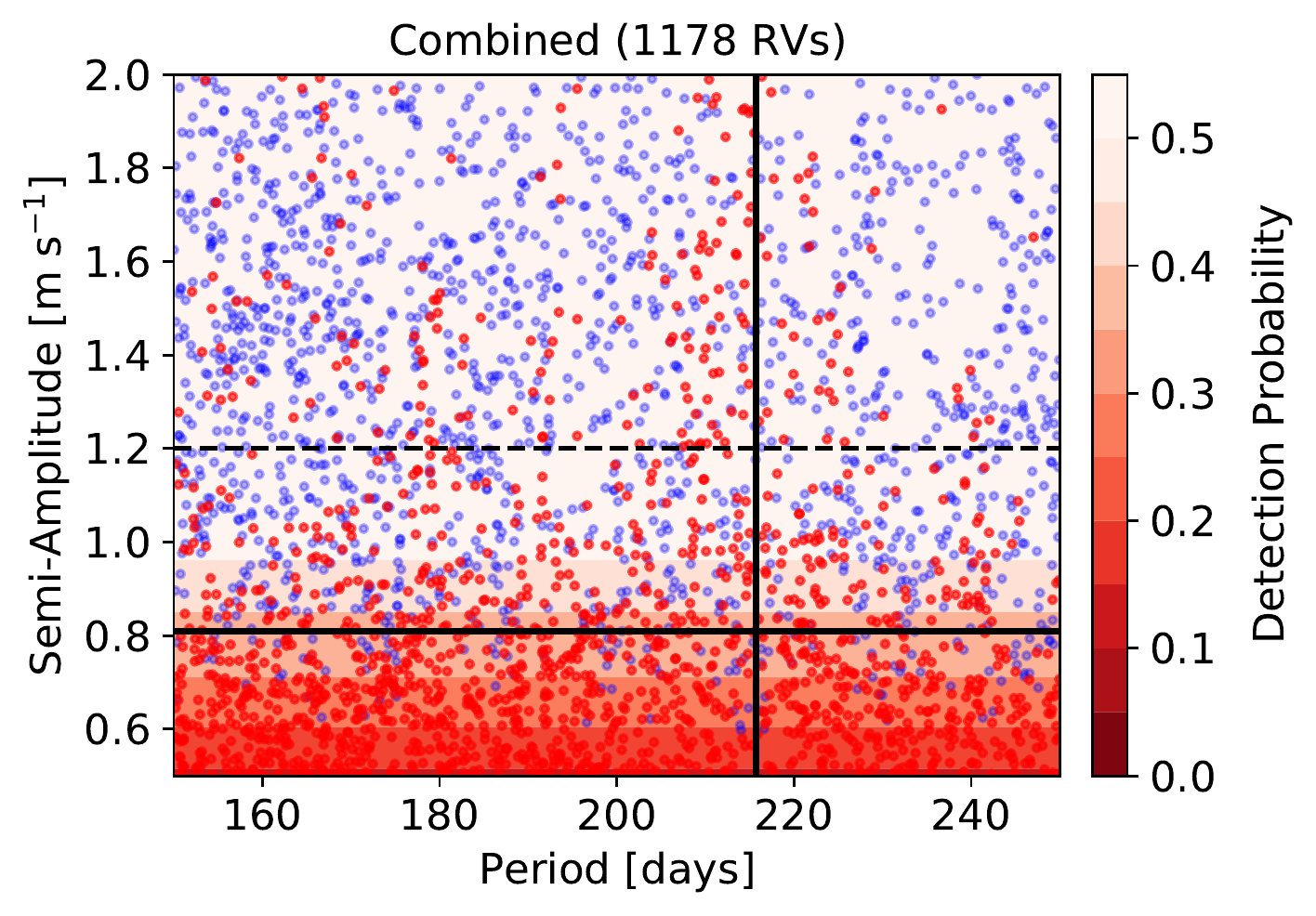}
    \caption{Completeness maps for the APF (top left), HIRES (top right), SOPHIE (middle left), CARMENES (middle right), and complete data sets (bottom). Signals that were recovered by {\tt RVSearch} are denoted by blue points while signals that were not detected are given by red points. The contours represent the probability that a planet with the corresponding period and amplitude is detected. The vertical black line marks the period of the candidate signal while the horizontal solid black line gives the amplitude from our adopted model (see \rfsecl{model}). The dashed horizontal black line marks the elevated amplitude from models that ignore the stellar rotation signal. For both amplitudes, HIRES and SOPHIE are unlikely to detect the candidate signal while the APF and combined RVs are more likely to make a detection than CARMENES.}
    \label{fig:injection}
\end{figure*}

\subsection{Investigating the Origin of the Candidate Signal}
\label{sec:systematic}

\begin{figure*}[!t]
    \centering
    \includegraphics[width=\linewidth]{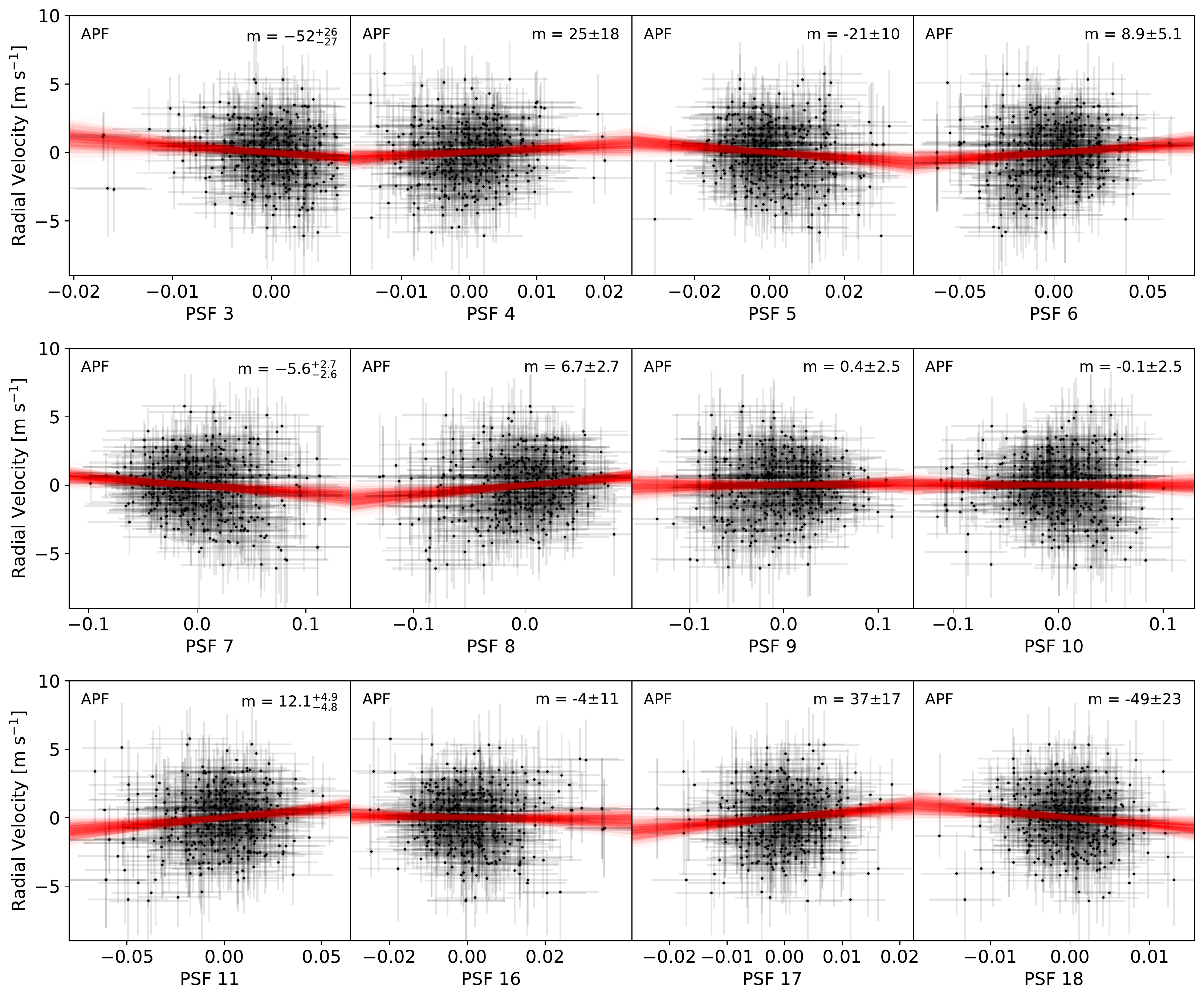}
    \caption{Plots of the radial velocities versus PSF parameters for the APF data set. Signals corresponding to GJ 411 b, GJ 411 c, and stellar rotation have been subtracted from the RVs using the model presented in \rfsecl{model}. Similar to \rffigl{actcorr}, samples of linear models from the MCMC posterior for each pair of variables are shown as red lines while the $68.3\%$ confidence interval for the slope is given in the top-right corner of each subplot. Most of the PSF parameters show some correlation with the RVs; however they appear to be weak and are not strictly caused by the candidate signal.}
    \label{fig:psfcorr}
\end{figure*}

\begin{figure*}[!t]
    \centering
    \includegraphics[width=\linewidth]{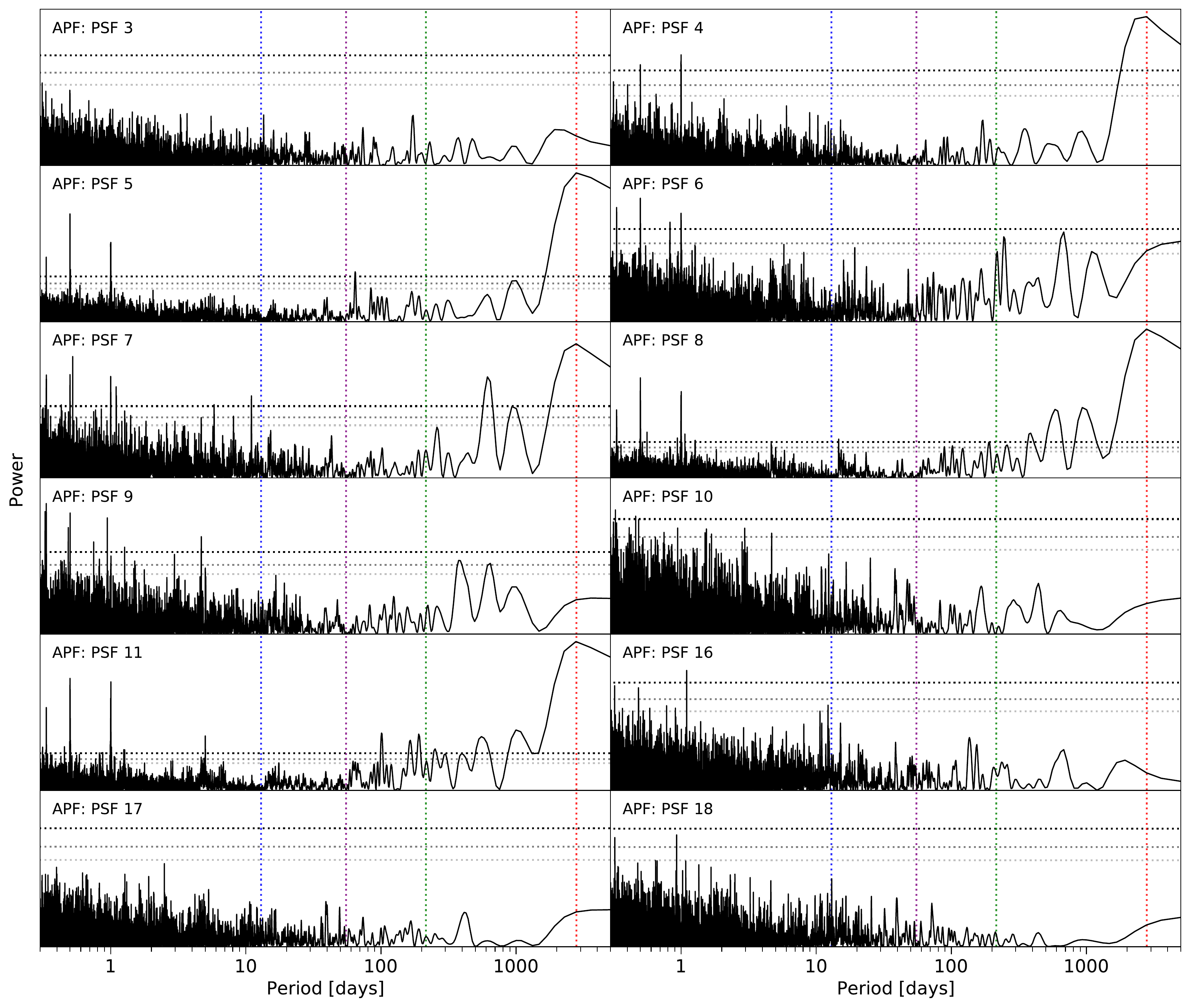}
    \caption{GLS periodograms of the APF PSF parameters. As with \rffigl{actper}, the three horizontal dashed lines in each panel represent the $1\%$, $0.05\%$, and $0.01\%$ FAP levels for the corresponding periodogram. Additionally, the vertical dashed lines represent the periods of interest, with blue marking GJ 411 b, purple giving the rotational modulation, green denoting the candidate signal, and red representing GJ 411 c. No peak near 215 days appears for any of the PSF parameters, suggesting that the candidate signal is not driven by a systematic.}
    \label{fig:psfper}
\end{figure*}

\cite{Rosenthal:2021} detected a $214.59\pm0.82$ day signal in a joint analysis of their APF and HIRES data. This signal was primarily dismissed as a systematic because it was prominent within the APF RVs but went undetected in the HIRES data. As shown in \rffigl{periodograms} and discussed in \rfsecl{discovery}, our HIRES periodogram does show a peak at 215 days; however it does not fall above the $0.001$ FAP threshold and is not flagged by the {\tt RVSearch} algorithm. The CARMENES and SOPHIE periodograms do not show any power near this period; however, the peak at 215 days appears to be more prominent in the combined data set than in the APF periodogram with a FAP of $2.56\times10^{-10}$ versus $1.38\times10^{-6}$. If we only include the HIRES or CARMENES RVs with the APF RVs, the FAP of the 215 day periodicity still improves, suggesting that the signal is present in each of these three data sets. Combining the SOPHIE RVs with the APF data marginally decreases the peak's FAP; however this is likely because the rotation signal at 55 days becomes much stronger. Visually, the 215 day peak continues to rise further above the noise floor in the periodogram.

To determine whether the signal could be present but go undetected in three of our four RV data sets, we perform injection recovery tests. For each instrument, we randomly generate 3000 signals with periods drawn from a uniform distribution between 150 and 250 days, amplitudes drawn from a uniform distribution between $0.5$ and $2$ \ms, eccentricities drawn from a beta distribution described by \cite{Kipping:2013}, and arguments of periastron and times of periastron passage drawn from uniform distributions. Each Keplerian is then separately injected into our RVs, and if {\tt RVSearch} is able to recover a significant peak near the expected period, we consider the signal detectable. Using our results, we create completeness maps revealing the detection limits \citep{Howard:2016} for each instrument, shown in \rffigl{injection}. Our adopted model in \rfsecl{model} gives an amplitude of $\kdavgerr$ \ms, which corresponds to detection probabilities of nearly zero in the HIRES and SOPHIE data. The CARMENES radial velocities have a $\mysim7.5\%$ chance of detecting the signal whereas the APF RVs have an $\mysim11\%$ chance. If we ignore the rotation signal in our modeling, we find that the signal has a larger amplitude of $1.22\pm0.16$ \ms (see \rfsecl{model}); however the APF data set is still the most likely to detect the 215 day periodicity. It is feasible that the candidate signal is real and present in all four data sets despite only being detected by the APF.

\cite{Rosenthal:2021} also noted that the candidate signal had a high eccentricity, with their {\tt RVSearch} output giving values of $0.52\pm0.18$. This was used to further dismiss the signal as a systematic. In contrast, our modeling returns an eccentricity of $\eccdavgerr$, consistent with a circular orbit. We attribute this discrepancy to the stellar rotation signal; when fitting three-Keplerian models and ignoring rotational modulation, we also find elevated eccentricities for the candidate signal. It is likely that the stellar rotation period, which falls near a harmonic of 215 days, is interfering with the 215 day signal near its maxima and minima, driving artificially high amplitudes and eccentricities.

\begin{figure}[!t]
    \centering
    \includegraphics[width=\linewidth]{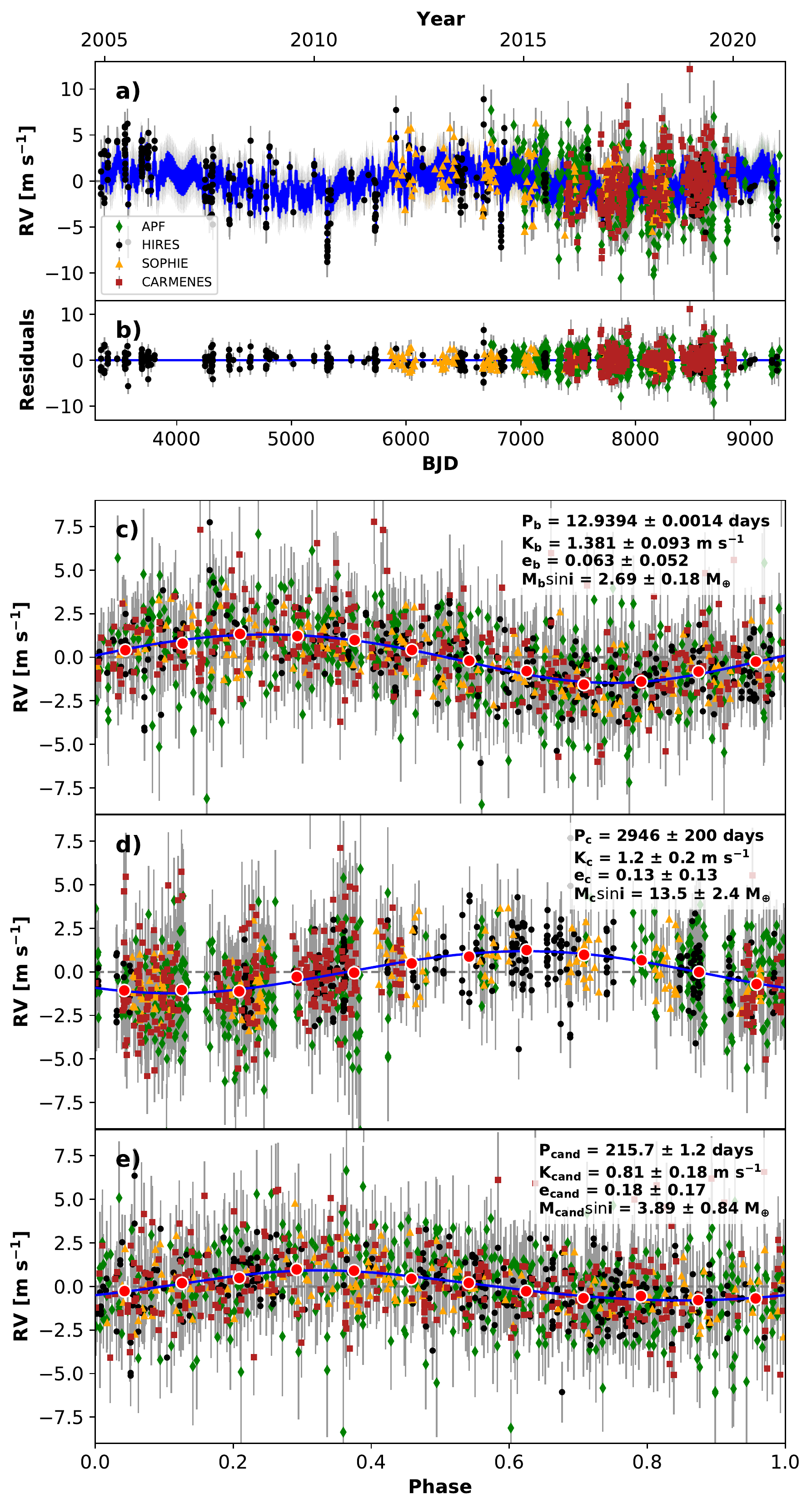}
    \caption{GJ 411 radial velocities with error bars representing observational uncertainties and instrumental jitter values added in quadrature. Panel a: Full time series with the best fit three-planet model and mean GP shown in blue. Panel b: residuals of the best fit model and mean GP. Panel c: RVs phase-folded to the ephemeris of GJ 411 b. The Keplerian orbits of all other planets and the rotation signal have been subtracted and the Keplerian fit for the individual planet is shown in blue. Red dots represent the binned velocities. Panels d and e are the same, except they represent the orbits of GJ 411 c and the candidate signal, respectively.}
    \label{fig:rv}
\end{figure}

Common sources of systematics such as instrumental hysteresis should affect the point spread function (PSF) parameters used to model the APF spectra in addition to the radial velocities. Following a procedure similar to that in \rfsecl{distinct}, we search for correlations between the APF RVs and each PSF parameter. This time, however, we remove the signals corresponding to GJ 411 b, GJ 411 c, and stellar rotation from the RVs, allowing us to primarily search for a 215 day systematic. Additionally, each of the PSF parameters is tightly clustered around some central value, but outliers continue to affect linear models of the data. We perform an initial three-sigma clip on each PSF parameter and then remove any points further than five median absolute deviations from the mean. In some cases, the PSF parameters show too little variation to fit a model and are excluded from our analysis. As seen in \rffigl{psfcorr}, nine out of twelve PSF parameters show correlation with the RVs; however the actual data have very large scatters. We also note that the results are highly dependent on the method used to reject outliers. GLS periodograms of the PSF parameters in \rffigl{psfper} show that none have significant power near 215 days, confirming that any correlations between the RVs and PSF parameters are not associated with the candidate signal. Ultimately, we find little evidence that a systematic drives the 215 day periodicity in the APF data. However, given that we cannot independently detect the signal in any of the other three data sets, we continue to consider it to be a candidate planet.

\subsection{Modeling Signals of Interest}
\label{sec:model}

\input{planetaryparams}

We model the three planetary and candidate signals with Keplerian orbits using the Python package {\tt exoplanet} \citep{dfm:2021}. Additionally, we choose to represent the rotational modulation with a Gaussian process (GP; \citealt{Rasmussen:2006, Roberts:2012}), which are known to represent the effects of active regions on the surface of a star \citep{Haywood:2014, Rajpaul:2015}. Further, GPs give us the flexibility needed to account for changes in rotational modulation between the four instruments, particularly as they cover different wavelengths. We use the rotation kernel implemented in {\tt celerite2} \citep{celerite1, celerite2}, which combines two dampened simple harmonic oscillators with periods of $P$ and $P/2$ to capture the stochastic variability in a star's rotational modulation. This GP takes five parameters, including the rotation period ($\rho$), the quality factor for the secondary mode ($Q_0$), the difference between the quality factors for the two modes ($dQ$), the amplitude ($\sigma$), and the fractional amplitude of the secondary mode compared to the primary ($f$).

No linear or quadratic terms are included in our model because they are highly covariant with the signal for GJ 411 c. However, we include jitter ($\xi$) and RV-offset ($\gamma$) terms for each instrument. We place uniform priors on several MCMC step parameters (period, eccentricity, argument of periastron, and jitter) to keep them within physical bounds. We additionally place a modified Jeffreys prior on the amplitude for each planet and the rotation signal \citep{Gregory:2007, Haywood:2014}; by choosing a knee equal to the mean estimated uncertainty of our RVs, we can ensure that the amplitudes are not overestimated in the case of a nondetection. The final log-likelihood for this model is

\begin{equation}
    \mathcal{L} = -\frac{1}{2}\mathbf{r_\theta}^T K_\alpha^{-1}\mathbf{r_\theta} - \frac{1}{2}\ln\det K_\alpha - \frac{N}{2}\ln(2\pi),
\end{equation}
where $\mathbf{r_\theta}$ is the vector of residuals from our mean function, calculated using the three Keplerians, $K_\alpha$ is the kernel function estimated using {\tt celerite2}, and $N$ is the number of data points.

Using the No-U-Turns algorithm (NUTS; \citealt{Hoffman:2011}) implemented through {\tt PyMC3} \citep{pymc3}, we run an MCMC process to explore the parameter space. Our analysis used 8 chains and was well-converged after $\numprint{16000}$ draws with a final maximum Gelman–Rubin statistic \citep{Gelman:1992} of 1.0005. The resulting posterior is shown in \rftabl{pparams} and the best-fit model in \rffigl{rv}.

\input{logps}

We consider a number of other models, including those that represent the rotational modulation as a simple Keplerian. We find that the Keplerian is an inadequate model given that the rotation signal is not strictly periodic over the baseline covered by our data and that it has different amplitudes in the four different data sets. Ignoring the rotation signal altogether returns parameters consistent with those presented in \rftabl{pparams} with the exception of elevated eccentricities ($0.41\pm0.15$ versus $\eccdavgerr$) and amplitudes ($1.22\pm0.17$ \ms\ versus $\kdavgerr$ \ms) for the candidate signal. As discussed in \rfsecl{systematic}, the rotation signal could possibly be constructively and destructively interfering with the candidate signal, artificially elevating both the eccentricity and amplitude. We caution that the reported uncertainties on $e_\mathrm{cand}$ and $K_\mathrm{cand}$ may be under-reported. We also fit a two-Keplerian model only representing the signals from GJ 411 b and c in addition to a GP representing the rotation. The results are consistent with the values given in \rftabl{pparams}, indicating that if the candidate signal is an APF systematic, it does not interfere with the orbital solutions for the confirmed planets. However, as shown in \rftabl{logps}, three-planet solutions are the best fit to each individual data set, providing further evidence in support of the candidate planet.

\section{Discussion}
\label{sec:discussion}

\subsection{Comparisons to Previous Work}
\label{sec:comparisons}

\input{compare}

By combining the APF, HIRES, SOPHIE, and CARMENES data, we have a more comprehensive data set than \cite{Diaz:2019}, \cite{Stock:2020}, or \cite{Rosenthal:2021}, allowing our modeling to place tighter constraints on each of the orbital parameters for GJ 411 b and c, as seen in \rftabl{compare}. For GJ 411 b, our orbital period is formally inconsistent with the models from \cite{Diaz:2019} and \cite{Stock:2020}, however, all other parameters for the planet are consistent, including the mass and semi-major axis. While our orbital period, eccentricity, and semi-major axis for GJ 411 c are in agreement with the model put forward by \cite{Rosenthal:2021}, we find a much lower minimum mass comparable to that of Uranus; with a greater number of observations, our analysis is better suited to constrain the RV amplitude and mass of this long-period planet. Finally, even though \cite{Rosenthal:2021} dismissed the 215 day signal as a systematic, they provide the respective amplitude and period from their modeling, which are also shown in \rftabl{compare}. Although our adopted model returns a much lower amplitude for the candidate signal, this once again is likely because \cite{Rosenthal:2021} ignore stellar rotation.

\cite{Butler:2017} identified a planetary candidate with a period of $9.8693\pm0.0016$ days using HIRES RVs, which has not been confirmed by any other studies. \cite{Diaz:2019} conducted an independent analysis of these HIRES observations but were unable to identify a similar signal. Notably, these two analyses used different models, with \cite{Butler:2017} decorrelating the radial velocities using Ca II lines. This discrepancy could cause different signals to be observed. However, \cite{Diaz:2019} found a peak located at the same period in the window function, suggesting that the signal is not astrophysical. While processed through a different pipeline, many of our HIRES radial velocities come from the same spectra---although our analysis excludes any of the pre-2004 data. As seen in \rffigl{periodograms}, there is a stand-alone peak in the HIRES periodogram located before 10 days, although it does not appear to fall near 9.9 days. As emphasized by previous works, the \cite{Butler:2017} signal is likely a false detection that can be attributed to systematics or modeling artifacts. 

\subsection{Prospects for Follow-up}
\label{sec:dual}

\subsubsection{Transit Photometry}
\label{sec:transit}

Using the analytic relationship between mass and radius given by \cite{Weiss:2014}, we calculate the expected radius distribution for each planet and candidate from our RV model posterior. To account for the mass–inclination degeneracy, we randomly assign orbital inclinations drawn from a uniform cosine distribution to each sample. This yields radii of $1.18^{+0.75}_{-0.18}$ \rearth, $6.99^{+4.46}_{-1.66}$ \rearth, and $1.83^{+1.15}_{-0.50}$ \rearth\ for planets b, c, and the candidate, respectively. With the radius posteriors and the equation
\begin{equation}
    P_\mathrm{transit}=\left(\frac{R_\star + R_p}{a}\right)\left(\frac{1+e\sin\omega}{1-e^2}\right),
\end{equation}
we calculate the respective geometric transit probabilities to be $\mysim2.26\%$, $\mysim0.07\%$, and $\mysim0.36\%$.

\begin{figure}[!t]
\centering\includegraphics[width=\linewidth]{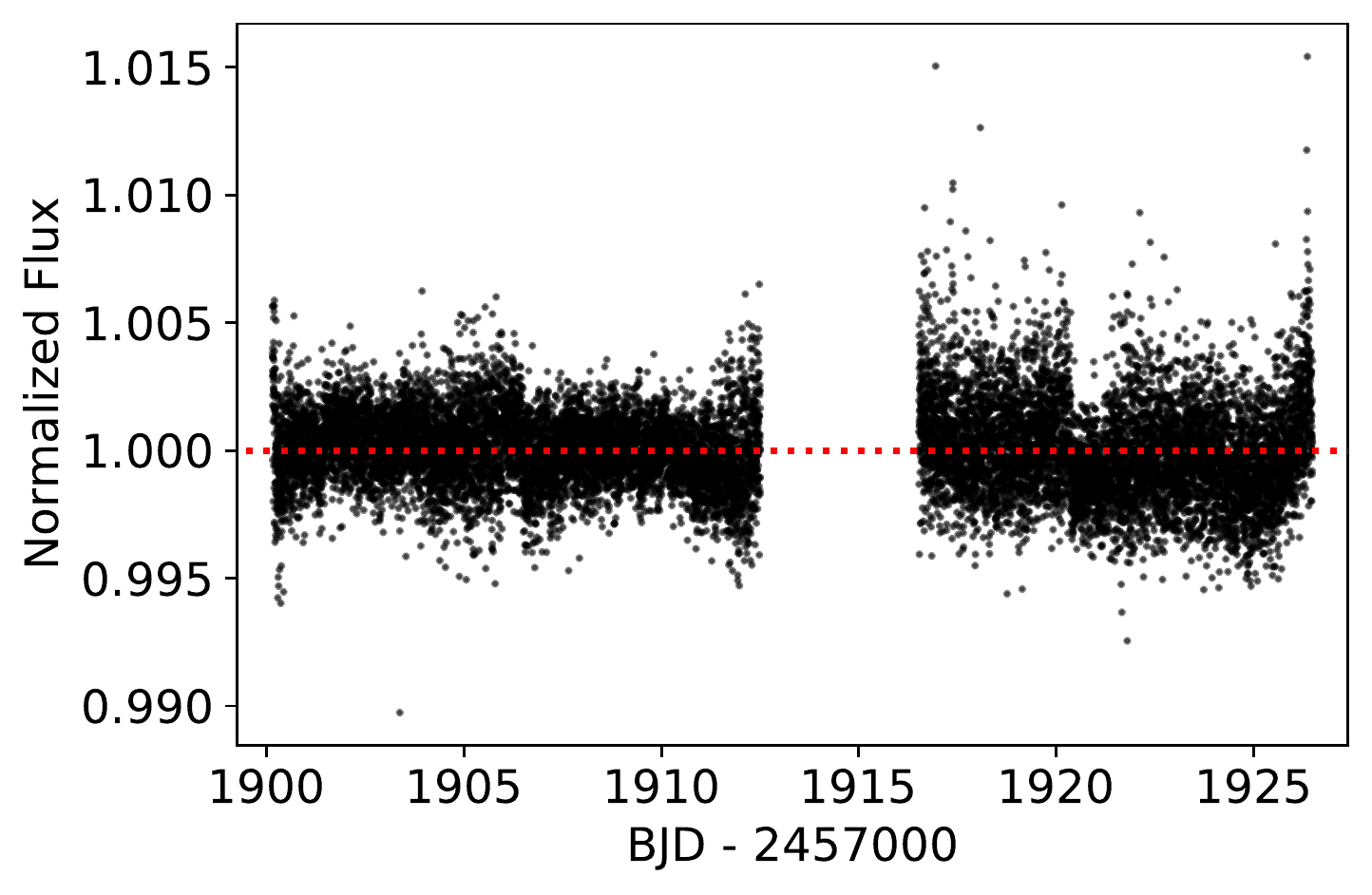}
\caption{The normalized PDCSAP \tess\ light curve for GJ 411 during sector 22. The horizontal red dashed line marks a constant normalized flux of 1, from which there is little deviation and no apparent transiting event.}
\label{fig:lightcurve}
\end{figure}

\begin{figure*}[!tb]
\centering\includegraphics[width=\linewidth]{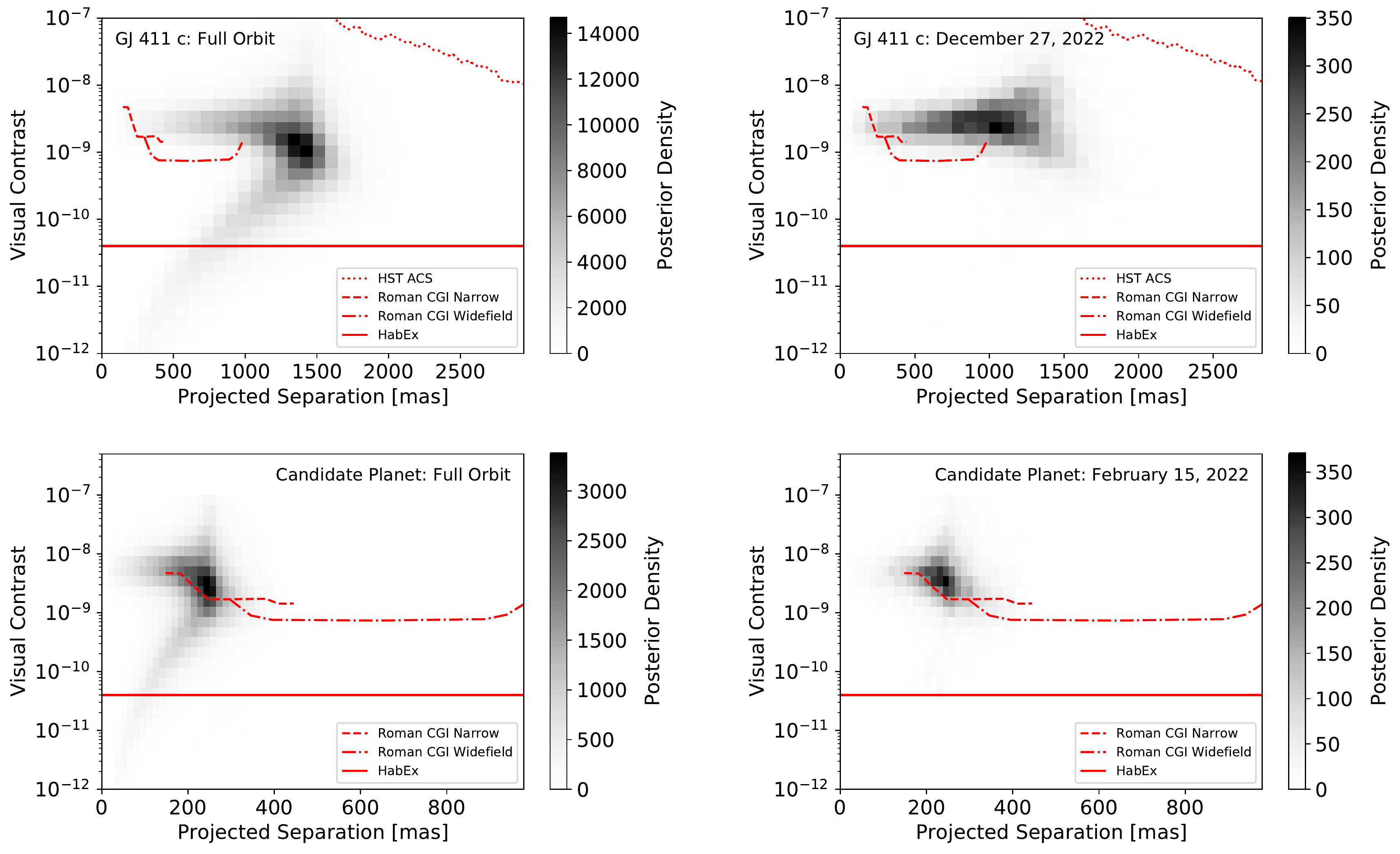}
\caption{Two-dimensional histograms showing the joint planet contrast and projected separation posteriors for GJ 411 c (top row) and the candidate planet (bottom row). Both are approximated as a Lambertian disk with a Bond albedo of 0.5. Panels on the left show the posterior for the entire orbit while panels on the right show the posterior for a single time where the planet would be most easily observed. GJ 411 c is expected to be most easily observed on December 22, 2023 while the candidate would be best detected on February 15, 2022. The predicted post-processing detectability floor is shown for the Roman Space Telescope CGI in narrow and wide field-of-view modes, along with the HabEx performance requirement. We also show the detection limits for HST ACS in the upper row. Current instrumentation, such as HST ACS, does not have the sensitivity needed to detect reflected light from GJ 411 c and the candidate planet, but future missions such as Roman or HabEx may be able to image the planets in the visual spectrum.}
\label{fig:reflected}
\end{figure*}

We then calculate the fraction of orbits in our posterior with transit times falling between the start and end of \tess\ sector 22. Multiplying these by the geometric probabilities, we find that the probabilities of GJ 411 b, c, and the candidate transiting during our \tess\ data are $\mysim2.25\%$, $<6\times10^{-3}\%$, and $\mysim0.06\%$, respectively.

As seen in \rffigl{lightcurve}, there is no visible transit in the PDCSAP \tess\ data. However, we search the light curve using the package {\tt Transit Least Squares} ({\tt TLS}; \citealt{Hippke:2019}) and confirm \cite{Stock:2020}'s findings that the strongest periodicity is located at $13.0$ days but has a negligible signal detection efficiency of 7.1. While no apparent transits have occurred during \tess\ sector 22, during which GJ 411 b would have occulted twice, it is possible that the signal is obscured by noise. Given the expected radius, we anticipate a transit depth of $\mysim1$ ppt; however the RMS of the light curve is $1.7$ ppt.

\subsubsection{Direct Imaging}
\label{sec:imaging}

\begin{figure*}
\centering\includegraphics[width=\linewidth]{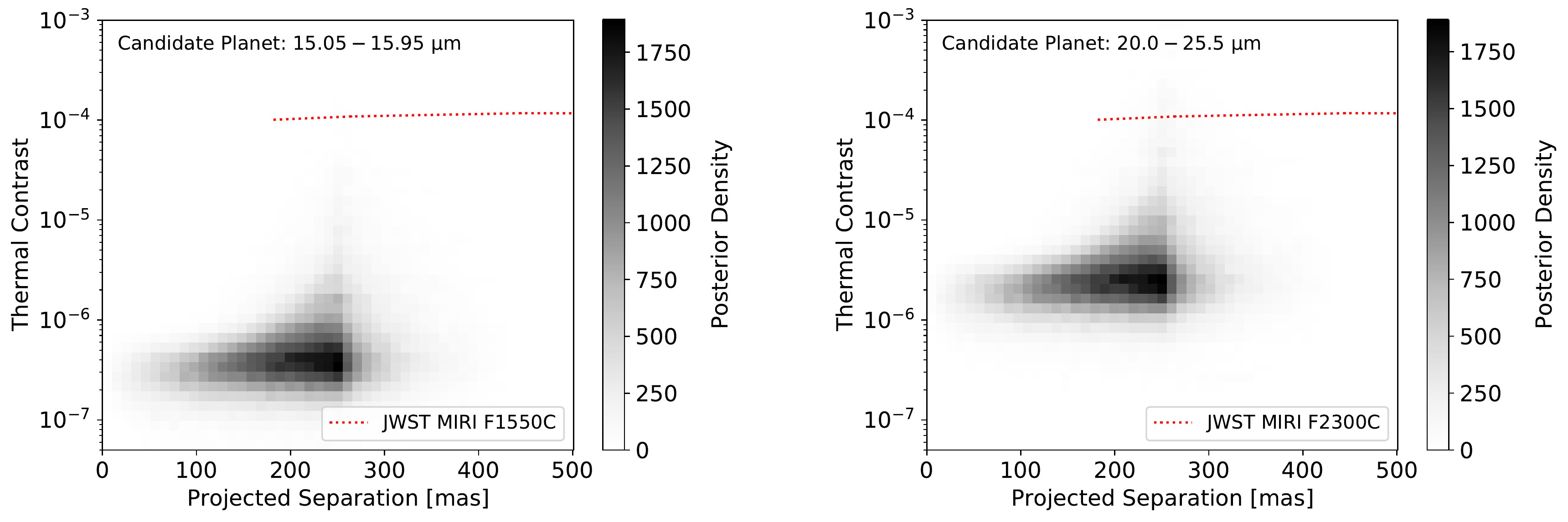}
\caption{Two-dimensional histograms showing the detectability of the candidate planet in the thermal infrared, assuming that the planet and star are perfect black bodies and the orbit is circular. The left plot shows the contrast through the JWST MIRI F1550C filter while the right shows the contrast through the JWST MIRI F2300C filter. The red lines represent the limiting contrast for the coronagraph \citep{Boccaletti:2015}. The candidate is likely too dim for MIRI to image.}
\label{fig:thermal}
\end{figure*}

Orbiting a close, relatively bright M dwarf, GJ 411 c and the candidate planet are prime targets for imaging. To explore the prospects for a direct detection, we follow a procedure similar to that outlined by \cite{Blunt:2019}. We derive the projected separation posterior for the outermost planet and candidate at future epochs throughout their orbits with the expression 
\begin{equation}
    \Delta\theta = \frac{r}{d}\left[\cos^2(\omega+f) + \sin^2(\omega+f)\cos^2(i)\right]^{1/2}
\end{equation}
\citep{Kane:2011}. Here, $r$ is the star–planet separation, $d$ is the star–observer distance, $\omega$ is the argument of periastron, $f$ is the true anomaly, and $i$ is the orbital inclination randomly assigned in \rfsecl{transit}.

\begin{figure}[!b]
\centering\includegraphics[width=\linewidth]{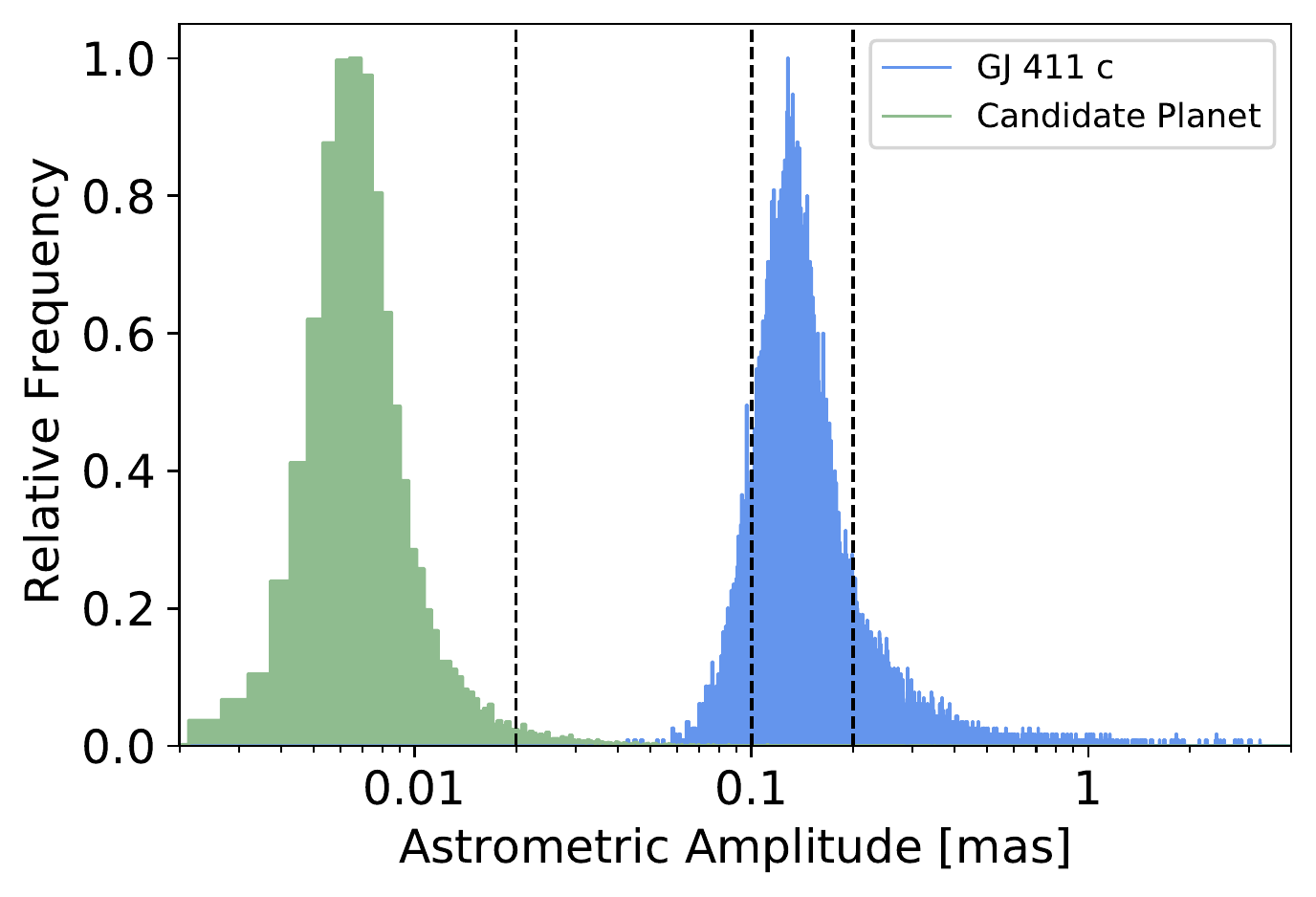}
\caption{Histograms showing the astrometric amplitude posterior for GJ 411 c (blue) and the candidate (green). Assuming Gaia has uncertainties of 0.02 mas, the vertical dashed lines mark one-sigma, five-sigma, and ten-sigma detection limits. For an orbit to be detectable, we require the astrometric amplitude to be greater than ten-sigma. $97\%$ and $2\%$ of simulated orbits for GJ 411 c and the candidate planet, respectively, are detectable with Gaia.}
\label{fig:astrometric}
\end{figure}

We then estimate the planets' expected reflected visual light contrasts, approximating both as Lambertian disks with scattering albedos of $0.5$---roughly between the albedos of Earth ($\mysim0.3$) and Venus ($\mysim0.8$). The phase angle, $\alpha$, is calculated at each epoch for every sample in our posterior using the relation
\begin{equation}
    \cos(\alpha) = \sin(f+\omega)\sin(i).
\end{equation}
With the radii posteriors derived in \rfsecl{transit}, the expected contrast from the star is evaluated using the equation
\begin{equation}
    \frac{F_p}{F_\star} = A_g\left(\frac{R_p}{a}\right)^2\Phi(\alpha),
\end{equation}
where the geometric albedo is related to the scattering albedo by
\begin{equation}
    A_g = \frac{2}{3}A_s
\end{equation}
and 
\begin{equation}
    \Phi(\alpha) = \frac{\sin(\alpha) + (\pi-\alpha)\cos(\alpha)}{\pi}
\end{equation}
\citep{Madhusudhan:2012}. \rffigl{reflected} shows two-dimensional histograms comparing the planet–star contrast distribution to the projected separation distribution, along with the expected detection limits for several instruments. While GJ 411 c and the candidate are not detectable by instrumentation such as the HST Advanced Camera for Surveys (ACS), which represents some of the best detection limits currently available, future space missions such as the Nancy Grace Roman Telescope will likely be able to detect both at certain points in their orbits. Rough estimates of HabEx performance indicate that future missions may also be able to detect them at most times\footnote{Anticipated detection limits obtained from \href{https://github.com/nasavbailey/DI-flux-ratio-plot}{github.com/nasavbailey/DI-flux-ratio-plot}}.

To determine whether the candidate planet could be imaged in the infrared, we derive its equilibrium temperature posterior, assuming a Bond albedo of $0.5$ and a circular orbit; although the 215 day signal is somewhat eccentric, variations in the equilibrium temperature due to orbital position are negligible. Assuming both the star and planet behave as ideal blackbodies, we take the ratios of their fluxes and integrate them over different bandpasses. The resulting posteriors are shown for two JWST Mid-Infrared Instrument (MIRI) filters, along with the instrument's limiting contrast curve \citep{Boccaletti:2015}, in \rffig{thermal}. We see that the planet's flux is consistently too low to be captured by MIRI.

\subsubsection{Astrometry}
\label{sec:astrometry}

Given the star's proximity, GJ 411 is also a prime target for astrometry. After using the inclinations drawn in \rfsecl{transit} to account for the mass–inclination degeneracy, we calculate the expected astrometric amplitude for each planet using the equation
\begin{equation}
    A_\theta = \frac{M_p}{M_\star + M_p}\frac{a}{d}.
\end{equation}
The resulting distributions for GJ 411 c and the candidate planet are shown in \rffigl{astrometric}. Assuming that each planet's signal is distinguishable from proper motion, we consider an orbit to be detectable if its amplitude is five times greater than the expected uncertainty. \cite{Gaia:2020b} gave a typical uncertainty of $0.02$ mas during a single epoch for stars with magnitudes below 15. 
By this standard, $92\%$ of GJ 411 c's and $2\%$ of the candidate's simulated orbits are detectable. While the $\mysim7.5$ mag star would likely have greater uncertainties, multiple measurements would contribute to a more significant detection.

Gaia Data Releases 1 \citep{Gaia:2016B,Gaia:2016A} and 2 \citep{Gaia:2018A} did not contain any observations of Gliese 411. The significance of the excess noise can indicate whether an astrometric solution behaves well, with values greater than 2 indicating a poor fit. In Early Data Release 3 \citep{Gaia:2020b}, this value for GJ 411 was 36.9, with the caveat that modeling errors may inflate the significance of the excess noise for an early data release. In the case that the astrometric model poorly fits Gaia data from GJ 411, this would suggest the presence of detectable astrometric signals.

\subsection{Comparative Exoplanetology}
\label{sec:exoplanetology}

Only 2.55 pc away from the solar system, GJ 411 is a remarkably nearby star, with only the Alpha Centauri system, Barnard's Star, and Wolf 359 closer. While each of these stars or systems has at least one planet candidate \citep{Anglada-Escud:2016,Ribas:2018,Tuomi:2019,Damasso:2020}, GJ 411 could potentially be the closest star with three or more planets.

Assuming circular orbits and Bond albedos of 0.5, the equilibrium temperatures for GJ 411 b, c, and the candidate---derived from the RV model posterior---are $\mysim326$ K, $\mysim53$ K, and $\mysim128$ K, respectively. Using the conservative definition of the habitable zone in \cite{Kopparapu:2013}, where the inner edge is given by the moist greenhouse model and the outer edge by the maximum greenhouse limit, we find that the habitable zone extends from $0.14\pm0.007$ au to $0.258\pm0.012$ au. Planet b and the candidate both straddle this region but never enter it throughout their full orbits. 

\begin{figure}[!t]
\centering\includegraphics[width=\linewidth]{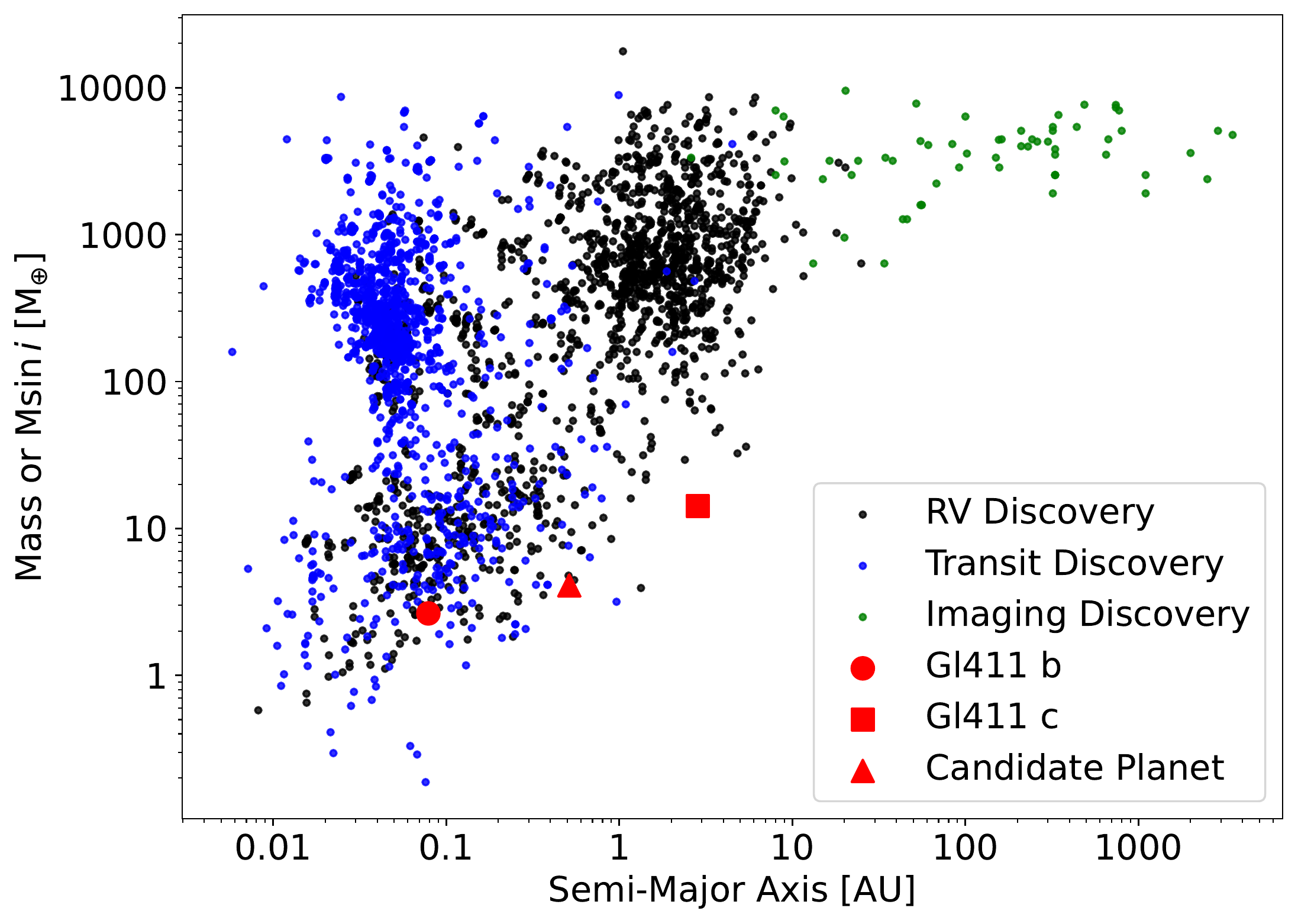}
\caption{Mass or M$\sin{i}$ versus semi-major axis for planets in the NASA Exoplanet Archive (Retrieved 2021 May 2) and the GJ 411 system. GJ 411 b, c, and the candidate have relatively low minimum masses compared to other known planets at similar orbital distances. Additionally, despite GJ 411 c and the candidate possibly being imageable, they have much smaller semi-major axes and masses than planets that have been discovered using direct imaging.} 
\label{fig:population}
\end{figure}

While the majority of planets detected through imaging have masses on the scale of thousands of \mearth, recent observations of Proxima c using SPHERE at the Very Large Telescope \citep{Gratton:2020} show that the direct detection of small, nearby planets---particularly those around bright stars---may not be far out of reach. Even though the Roman telescope and missions similar to the HabEx design promise to image a smaller class of planets than previously possible, GJ 411 c and the candidate planet would still be unusually small and close to their star for a direct image. Consequently, the Neptune and potential super-Earth could be among the first imaged planets of their kind. \rffigl{population} illustrates how different both are from objects that are conventionally imaged. 

It is also worth noting that GJ 411 does not show large RV variations and each signal of interest has a low amplitude (<1.5\ms). At the time of its discovery, GJ 411 b was the lowest-amplitude signal detected by SOPHIE; however, the instrument has since aided in the confirmation of HD 158259 b, which has an amplitude of $\mysim1.05$ \ms\ (although the actual discovery was made using \tess\ data; \citealt{Hara:2020}). The candidate signal at 215 days has an incredibly low amplitude of $\kdavgerr$ \ms, although this value could be somewhat larger due to limitations in our modeling. While HIRES data were used in the detection of the candidates around $\tau$ Ceti, which do have lower amplitudes \citep{Tuomi:2013, Feng:2017}, this signal, if confirmed, could be one of the most sensitive detections made primarily using HIRES and APF observations. Regardless, GJ 411 c is the next lowest-amplitude signal ($\kcavgerr$) discovered with these instruments and is followed by HD 7924 d ($\mysim1.65$ \ms; \citealt{Fulton:2015}).

\section{Conclusions}
\label{sec:conclusions}

We perform a joint analysis of APF, HIRES, SOPHIE, and CARMENES radial velocities from the bright, nearby M dwarf Gliese 411, rexdetecting 12.9, 2900, 215, and 55 day periodicities. The first of these signals corresponds to the previously detected GJ 411 b. The 2900 day signal has been contested by previous works as either a planet or long-period stellar magnetic cycles. We analyze stellar activity metrics including $\mathrm{logR'_{HK}}$ and $\mathrm{H_\alpha}$ indices to conclude that this disputed signal corresponds to a planet, GJ 411 c. The 215 day signal was originally identified as an APF systematic, not showing in periodograms for other instruments. However, injection recovery tests show that the signal could have been obscured by cadence and noise in the other data sets, accounting for the nondetections. Further, the signal becomes more significant when combining RVs from all four sources and three-planet models best describe each individual data set. In an analysis of the PSF parameters used to fit the APF spectra, no signs of a 215 day systematic are found. While we cannot conclusively determine the cause of this signal, we consider the planetary interpretation to be most probable. Lastly, the 55 day periodicity most likely corresponds to rotational modulation detected in previous photometric work.

We model the three planetary and candidate signals with Keplerians and represent the stellar rotation with a Gaussian process. Additional fits ignoring the rotational modulation do not significantly alter orbital parameters except for the candidate amplitude and eccentricity. The rotation signal, located near a harmonic of 215 days, likely interferes with the candidate signal, accounting for these discrepancies. Additionally, two-Keplerian models representing GJ 411 b and c are consistent with the three-Keplerian model, verifying that our orbital solutions are accurate even in the case where the 215 day signal does not correspond to a planet. Our adopted model generally agrees with previous models for GJ 411 b and c while providing tighter constraints on their masses and orbital radii. The candidate signal would correspond to a super-Earth falling between the other planets with a minimum mass of $\msinidavgerr$ \mearth\ and a semi-major axis of $\adavgerr$ au.

With two confirmed planets and potentially three or more orbital companions, GJ 411 is one of the closest multiplanet systems to ours. While a secondary detection through transit photometry is not promising, direct imaging with missions such as the Nancy Grace Roman Telescope offers the chance to make a dual detection of GJ 411 c and to confirm the candidate planet. Additionally, astrometry through Gaia could potentially make a secondary detection of GJ 411 c. While similar observations are not possible for GJ 411 b---the innermost planet---in the near future, learning more about the architecture of the outer system is important for understanding its formation and evolution. Most importantly, this system could offer the opportunity to directly image lower-mass planets. 

\acknowledgements
We thank Michael Meyer for feedback and insight into determining a planet's detectability in the infrared.
This research has made use of the NASA Exoplanet Archive, which is operated by the California Institute of Technology, under contract with the National Aeronautics and Space Administration under the Exoplanet Exploration Program.
This work made use of the SIMBAD database (operated at CDS, Strasbourg, France), NASA’s Astrophysics Data System Bibliographic Services.
This research has made use of the VizieR catalog access tool, CDS, Strasbourg, France (DOI: 10.26093/cds/vizier). The original description of the VizieR service was published in A$\&$AS 143, 23.
This work has made use of data from the European Space Agency (ESA) mission {\it Gaia} (\url{https://www.cosmos.esa.int/gaia}), processed by the {\it Gaia} Data Processing and Analysis Consortium (DPAC, \url{https://www.cosmos.esa.int/web/gaia/dpac/consortium}). Funding for the DPAC has been provided by national institutions, in particular the institutions participating in the {\it Gaia} Multilateral Agreement.
L.M.W. is supported by the Beatrice Watson Parrent Fellowship and NASA ADAP Grant 80NSSC19K0597.
We are very grateful for the donations of the Levy family that helped facilitate the construction of the Levy spectrograph on the APF. Without their support, the APF would not be contributing to the discovery of planets like these.
Research at the Lick Observatory is partially supported by a generous gift from Google.
We are grateful to the time assignment committees of the University of Hawaii, the University of California, and NASA for their generous allocations of observing time. Without their long-term commitment to RV monitoring, these planets would likely remain unknown. We acknowledge R. Paul Butler and S. S. Vogt for many years of contributing to the data presented here. 
Finally, the authors wish to extend special thanks to those of Hawaiian ancestry on whose sacred mountain of Maunakea we are privileged to be guests. Without their generous hospitality, the Keck observations presented herein would not have been possible. 
\software{{\tt astropy} \citep{astropy:2018}, {\tt emcee} \citep{ForemanMackey:2012}, {\tt exoplanet} \citep{exoplanet:2021}, {\tt matplotlib} \citep{Hunter:2007}, {\tt numpy} \citep{Harris:2020}, {\tt RadVel} \citep{fulton:2018a}, {\tt RVSearch} \citep{Rosenthal:2021}}

\bibliographystyle{aasjournals}

\bibliography{refs}

\end{document}

%% file: newcommands.tex
\newcommand{\teff}{\ensuremath{T_{\rm eff}}\xspace}

\newcommand\vsini{\ifmmode{v\sin{i_\star}}\else $v\sin{i_\star}$\fi}
\newcommand\sini{\ifmmode{\sin{i_\star}}\else $\sin{i_\star}$\fi}

\newcommand{\msun}{\ensuremath{\,M_\Sun}}
\newcommand{\rsun}{\ensuremath{\,R_\Sun}}

\newcommand{\mjup}{\ensuremath{\,M_{\rm J}}}

\newcommand{\rearth}{\ensuremath{\,R_{\rm \Earth}}\xspace}
\newcommand{\mearth}{\ensuremath{\,M_{\rm \Earth}}\xspace}

\newcommand{\tess}{{\it TESS}}

\newcommand\mysim{\mathord{\sim}}

\newcommand{\ms}{\,m\,s$^{-1}$}

\newcommand{\rffig}[1]{Fig.~\ref{fig:#1}}
\newcommand{\rffigl}[1]{Figure~\ref{fig:#1}}

\newcommand{\rfsecl}[1]{\mbox{Section \ref{sec:#1}}}

\newcommand{\rftabl}[1]{Table~\ref{tab:#1}}

%% file: variables.tex
\newcommand{\perb}{12.9394^{+0.0014}_{-0.0013}}

\newcommand{\perc}{2946^{+210}_{-180}}

\newcommand{\perd}{215.7\pm1.2}

\newcommand{\tcb}{2456305.38^{+0.3}_{-0.28}}

\newcommand{\tcc}{2457205^{+98}_{-96}}

\newcommand{\tcd}{2456353^{+13}_{-16}}

\newcommand{\kb}{1.381^{+0.093}_{-0.092}}

\newcommand{\kc}{1.16^{+0.2}_{-0.19}}
\newcommand{\kcavgerr}{1.16\pm0.2}
\newcommand{\kd}{0.81^{+0.18}_{-0.17}}
\newcommand{\kdavgerr}{0.81\pm0.18}
\newcommand{\secoswb}{-0.15^{+0.2}_{-0.14}}

\newcommand{\secoswc}{-0.02^{+0.23}_{-0.24}}

\newcommand{\secoswd}{0.16^{+0.33}_{-0.35}}

\newcommand{\sesinwb}{-0.01^{+0.19}_{-0.18}}

\newcommand{\sesinwc}{0.15^{+0.31}_{-0.36}}

\newcommand{\sesinwd}{-0.05^{+0.35}_{-0.3}}

\newcommand{\eccb}{0.063^{+0.061}_{-0.043}}

\newcommand{\eccc}{0.132^{+0.16}_{-0.091}}

\newcommand{\eccd}{0.18^{+0.22}_{-0.13}}
\newcommand{\eccdavgerr}{0.18\pm0.18}
\newcommand{\omegab}{-0.5^{+3.2}_{-2.1}}

\newcommand{\omegac}{1.1^{+1.0}_{-3.1}}

\newcommand{\omegad}{-0.2^{+1.8}_{-1.4}}

\newcommand{\gpperiod}{55.17^{+0.27}_{-0.25}}

\newcommand{\gplogQ}{0.19^{+0.28}_{-0.14}}

\newcommand{\gplogdQ}{-1.3^{+1.2}_{-1.5}}

\newcommand{\gpf}{0.225^{+0.18}_{-0.088}}

\newcommand{\gpsigmaa}{1.55^{+0.21}_{-0.19}}

\newcommand{\gpsigmah}{2.13^{+0.27}_{-0.24}}

\newcommand{\gpsigmas}{1.8\pm0.26}

\newcommand{\gpsigmac}{1.36^{+0.27}_{-0.26}}

\newcommand{\gammaa}{-0.65^{+0.21}_{-0.2}}

\newcommand{\gammah}{-0.8\pm0.26}

\newcommand{\gammas}{-0.1^{+0.24}_{-0.23}}

\newcommand{\gammac}{0.25^{+0.25}_{-0.23}}

\newcommand{\sigmaa}{0.2^{+0.21}_{-0.14}}

\newcommand{\sigmah}{1.31^{+0.11}_{-0.1}}

\newcommand{\sigmas}{0.78^{+0.25}_{-0.29}}

\newcommand{\sigmac}{1.11\pm0.15}

\newcommand{\msinib}{2.69^{+0.19}_{-0.18}}

\newcommand{\msinic}{13.6^{+2.4}_{-2.3}}

\newcommand{\msinid}{3.89^{+0.82}_{-0.85}}
\newcommand{\msinidavgerr}{3.89\pm0.84}
\newcommand{\ab}{0.07879^{+0.00056}_{-0.00055}}

\newcommand{\ac}{2.94^{+0.14}_{-0.12}}

\newcommand{\ad}{0.5142^{+0.0043}_{-0.0041}}
\newcommand{\adavgerr}{0.5142\pm0.0042}
\newcommand{\tpb}{2456307.8^{+2.2}_{-4.8}}

\newcommand{\tpc}{2457250^{+820}_{-690}}

\newcommand{\tpd}{2456324^{+89}_{-46}}

\newcommand{\gplogpzero}{-3043.86}

\newcommand{\gplogpzerogp}{-2748.33}

\newcommand{\gplogponegp}{-2655.24}

\newcommand{\gplogptwogp}{-2630.5}

\newcommand{\gplogpthreegp}{-2614.77}

\newcommand{\planetslogpone}{-2924.52}

\newcommand{\planetslogptwo}{-2867.21}

\newcommand{\planetslogpthree}{-2823.98}

\newcommand{\apflogpzero}{-1071.79}

\newcommand{\apflogpone}{-1047.83}

\newcommand{\apflogptwo}{-1014.77}

\newcommand{\apflogpthree}{-984.43}

\newcommand{\hireslogpzero}{-757.1}

\newcommand{\hireslogpone}{-733.74}

\newcommand{\hireslogptwo}{-703.92}

\newcommand{\hireslogpthree}{-669.25}

\newcommand{\sophielogpzero}{-360.27}

\newcommand{\sophielogpone}{-339.69}

\newcommand{\sophielogptwo}{-337.67}

\newcommand{\sophielogpthree}{-327.3}

\newcommand{\carmeneslogpzero}{-739.14}

\newcommand{\carmeneslogpone}{-716.86}

\newcommand{\carmeneslogptwo}{-684.63}

\newcommand{\carmeneslogpthree}{-669.49}

%% file: affiliations.tex
\newcommand{\nexsci}{NASA Exoplanet Science Institute/Caltech-IPAC, 1200 E. California Boulevard, Pasadena, CA 91125, USA}

\newcommand{\caltech}{Department of Astronomy, California Institute of Technology, 1200 E. California Blvd, Pasadena, CA 91125, USA}

\newcommand{\boulder}{Department of Astrophysical and Planetary Sciences, University of Colorado, Boulder, CO 80309, USA}

\newcommand{\berkeley}{Department of Astronomy, University of California Berkeley, Berkeley, CA 94720, USA}

\newcommand{\notredame}{Department of Physics, University of Notre Dame, Notre Dame, IN 46556, USA}

\newcommand{\ucla}{Department of Physics and Astronomy, University of California Los Angeles, Los Angeles, CA 90095, USA}

%% file: authors.tex

\author[0000-0002-6903-9080]{Spencer A. Hurt}
\affiliation{\boulder}

\author[0000-0003-3504-5316]{Benjamin Fulton}
\affiliation{\nexsci}

\author[0000-0002-0531-1073]{Howard Isaacson}
\affiliation{\berkeley}

\author[0000-0001-8391-5182]{Lee J. Rosenthal}
\affiliation{\caltech}

\author[0000-0001-8638-0320]{Andrew W. Howard}
\affiliation{\caltech}

\author[0000-0002-3725-3058]{Lauren M. Weiss}
\affiliation{\notredame}

\author[0000-0003-0967-2893]{Erik A. Petigura}
\affiliation{\ucla}

%% file: rvs.tex
\begin{deluxetable}{lccccc}[!t]
\tabletypesize{\small}
\tablecaption{Radial Velocities of Gliese 411\label{tab:rvs}}
\tablehead{\multicolumn{1}{l}{BJD} & \multicolumn{1}{c}{RV} & \multicolumn{1}{c}{Uncertainty} &
\multicolumn{1}{c}{$\mathrm{H_\alpha}$} & \multicolumn{1}{c}{logR$\mathrm{'_{HK}}$} &
\multicolumn{1}{c}{Instrument} \\ 
-2450000 & $\mathrm{m\,s^{-1}}$ & $\mathrm{m\,s^{-1}}$ & & }
\startdata
\smallskip\\[-3.5ex]
8955.7945 & -0.6962 & 1.5564 & 0.0917 & -5.5499 & APF \\
8955.8042 & -1.1892 & 1.4823 & 0.0909 & -5.4362 & APF \\
8955.8106 & 1.4112 & 1.5855 & 0.0910 & -5.5268 & APF  \\
9185.0451 & -2.3714 & 1.8003 & 0.0895 & -5.4849 & APF \\
9185.0502 & -2.4402 & 1.9729 & 0.0891 & -5.5193 & APF \\
9185.0553 & -0.0376 & 1.7513 & 0.0893 & -5.5403 & APF \\
9192.9958 & -3.4514 & 1.7958 & 0.0904 & -5.3846 & APF \\[0.8ex]
\enddata
\tablecomments{This table is available in its entirety in machine-readable form.}
\end{deluxetable}

%% file: stellarparams.tex
\begin{deluxetable}{lrrr}
\tabletypesize{\small}
\tablecaption{Stellar Parameters for Gliese 411\label{tab:sparams}}
\tablehead{\multicolumn{1}{l}{Parameter} & \multicolumn{1}{r}{Value} & \multicolumn{1}{r}{Units}  & \multicolumn{1}{r}{Source}}
\startdata
\smallskip\\[-3.5ex]
~~~R.A. & 11 03 19.43 & hh mm ss & (1) \\
~~~Dec. & +35 56 55.15 & dd mm ss & (1) \\
~~~$\mu$\textsubscript{$\alpha$} & $-580.057 \pm 0.026$ & mas yr\textsuperscript{-1} & (1) \\
~~~$\mu$\textsubscript{$\delta$} & $-4776.589 \pm 0.030$ & mas yr\textsuperscript{-1} & (1) \\
~~~Parallax & $392.753 \pm 0.032$ & mas & (1) \\
~~~Distance & $2.54613 \pm 0.00021$ & pc & (1) \\
~~~\teff & $3719^{+21}_{-31}$ & K & (2) \\
~~~Rotation Period & $56.16 \pm 0.27$ & days & (3) \\
~~~Mass & $0.3899^{+0.0106}_{-0.0061}$ & \msun & (2) \\
~~~Radius & $0.3685^{+0.0081}_{-0.0054}$ & \rsun & (2) \\
~~~Age & $8.047^{+3.958}_{-4.523}$ & Gyr & (2) \\
~~~$\left[\rm{Fe/H}\right]$ & $-0.3621^{+0.0872}_{-0.0687}$ & \nodata & (2) \\
~~~log $g$ & $4.895^{+0.008}_{-0.010}$ & cgs & (2) \\
~~~$\mathrm{logR'_{HK}}$ & $-5.47^{+0.11}_{-0.09}$ & \nodata & (2) \\[0.8ex]
\enddata
\tablecomments{(1) Gaia Early Data Release 3 \citep{Gaia:2016B, Gaia:2020b}; (2) This Work; (3) \cite{Diaz:2019}}
\end{deluxetable}


%% file: planetaryparams.tex
\begin{deluxetable}{lrr}[p]
\tabletypesize{\small}
\tablecaption{Model Parameters and Derived Planetary Properties\label{tab:pparams}}
\tablehead{\multicolumn{1}{l}{Parameter} & \multicolumn{1}{r}{~~~~~~~~~~~~~~~~~~~~~~~~~~~~~~~Value} & \multicolumn{1}{r}{~~~~~~~~~~~~~~~~~~~~~~Units}}
\startdata
\smallskip\\[-3.5ex]
$P_b$ & $\perb$ & days \\
$T\rm{conj}_{\textit{b}}$ & $\tcb$ & JD \\
$\sqrt{e}\cos{\omega}_b$ & $\secoswb$ & \nodata \\
$\sqrt{e}\sin{\omega}_b$ & $\sesinwb$ & \nodata  \\
$K_b$ & $\kb$ & m s$^{-1}$ \\
$P_c$ & $\perc$ & days\\
$T\rm{conj}_{\textit{c}}$ & $\tcc$ & JD \\
$\sqrt{e}\cos{\omega}_c$ & $\secoswc$ & \nodata \\
$\sqrt{e}\sin{\omega}_c$ & $\sesinwc$ & \nodata \\
$K_c$ & $\kc$ & m s$^{-1}$ \\
$P_\mathrm{cand}$ & $\perd$ & days \\
$T\rm{conj}_{\mathrm{cand}}$ & $\tcd$ & JD \\
$\sqrt{e}\cos{\omega}_\mathrm{cand}$ & $\secoswd$ & \nodata \\
$\sqrt{e}\sin{\omega}_\mathrm{cand}$ & $\sesinwd$ & \nodata \\
$K_\mathrm{cand}$ & $\kd$ & m s$^{-1}$\\
$\rho$ & $\gpperiod$ & days \\
$\ln Q_0$ & $\gplogQ$ & \nodata \\
$\ln dQ$ & $\gplogdQ$ & \nodata \\
$f$ & $\gpf$ & \nodata \\
$\sigma$ (APF) & $\gpsigmaa$ & m s$^{-1}$ \\
$\sigma$ (HIRES) & $\gpsigmah$ & m s$^{-1}$ \\
$\sigma$ (SOPHIE) & $\gpsigmas$ & m s$^{-1}$ \\
$\sigma$ (CARMENES) & $\gpsigmac$ & m s$^{-1}$ \\
$\xi$ (APF) & $\sigmaa$ & m s$^{-1}$ \\
$\xi$ (HIRES) & $\sigmah$ & m s$^{-1}$ \\
$\xi$ (SOPHIE) & $\sigmas$ & m s$^{-1}$ \\
$\xi$ (CARMENES) & $\sigmac$ & m s$^{-1}$ \\
$\gamma$ (APF) & $\gammaa$ & m s$^{-1}$ \\
$\gamma$ (HIRES) & $\gammah$ & m s$^{-1}$ \\
$\gamma$ (SOPHIE) & $\gammas$ & m s$^{-1}$ \\
$\gamma$ (CARMENES) & $\gammac$ & m s$^{-1}$ \\
\midrule
$e_b$ & $\eccb$ & \nodata \\
$\omega_b$ & $\omegab$ & radians \\
$T\rm{peri}_{\textit{b}}$ & $\tpb$ & JD \\
$M_b \sin{i}$ & $\msinib$ & \mearth \\
$a_b$ & $\ab$ & au \\
$e_c$ & $\eccc$ & \nodata \\
$\omega_c$ & $\omegac$ & radians \\
$T\rm{peri}_{\textit{c}}$ & $\tpc$ & JD \\
$M_c \sin{i}$ & $\msinic$ & \mearth \\
$a_c$ & $\ac$ & au \\
$e_\mathrm{cand}$ & $\eccd$ & \nodata \\
$\omega_\mathrm{cand}$ & $\omegad$ & radians \\
$T\rm{peri}_\mathrm{cand}$ & $\tpd$ & JD \\
$M_\mathrm{cand} \sin{i}$ & $\msinid$ & \mearth \\
$a_\mathrm{cand}$ & $\ad$ & au \\[0.8ex]
\enddata
\tablecomments{Parameters correspond to the median MCMC chain values and uncertainties are given by the $68\%$ posterior credibility intervals.}
\end{deluxetable}

%% file: logps.tex
\begin{deluxetable*}{lrrrrr}[!t]
\tablecaption{Log-likelihoods for Different Models\label{tab:logps}}
\tablehead{\multicolumn{1}{l}{Model} & \multicolumn{1}{r}{~~~~~~~~~~~~~APF $\ln\mathcal{L}$} & \multicolumn{1}{r}{~~~~~~~~~~~~~~HIRES $\ln\mathcal{L}$} & \multicolumn{1}{r}{~~~~~~~~~~~~~~SOPHIE $\ln\mathcal{L}$} &
\multicolumn{1}{r}{~~~~~~~~~~~~~~CARMENES $\ln\mathcal{L}$} & \multicolumn{1}{r}{~~~~~~~~~~~~~~Combined $\ln\mathcal{L}$}}
\startdata
0 planets & $\apflogpzero$ & $\hireslogpzero$ & $\sophielogpzero$ & $\carmeneslogpzero$ & $\gplogpzero$ \\
1 planet & $\apflogpone$ & $\hireslogpone$ & $\sophielogpone$ & $\carmeneslogpone$ & $\planetslogpone$ \\
2 planets & $\apflogptwo$ & $\hireslogptwo$ & $\sophielogptwo$ & $\carmeneslogptwo$ & $\planetslogptwo$ \\
3 planets & $\apflogpthree$ & $\hireslogpthree$ & $\sophielogpthree$ & $\carmeneslogpthree$ & $\planetslogpthree$ \\
\midrule
0 planets + GP & \nodata & \nodata & \nodata & \nodata & \gplogpzerogp \\
1 planet + GP & \nodata & \nodata & \nodata & \nodata & \gplogponegp \\
2 planets + GP & \nodata & \nodata & \nodata & \nodata & \gplogptwogp \\
3 planets + GP & \nodata & \nodata & \nodata & \nodata & \gplogpthreegp \\
\enddata
\tablecomments{The orbital period and time of transit were fixed for GJ 411 c in the two- and three-planet models fit to the separate APF, HIRES, SOPHIE, and CARMENES data because each covers baselines too short to constrain the long-period signal.}
\end{deluxetable*}

%% file: compare.tex
\begin{deluxetable}{lrrr}[!b]
\tabletypesize{\small}
\tablecaption{Comparisons to Previous Work\label{tab:compare}}
\tablehead{\colhead{Parameter} & 
           \colhead{~~~Comparison Model} & 
           \colhead{~~~This Work} &
           \colhead{~~~Units}}
\startdata
\smallskip\\[-3.5ex]
\multicolumn{2}{l}{\cite{Diaz:2019}} &  & \\
~~~$M_b \sin{i}$ & $2.99 \pm 0.46$ & $\msinib$ & \mearth \\
~~~$a_b$ & $0.0785 \pm 0.0027$ & $\ab$ & au \\
~~~$P_b$ & $12.9532\pm0.0079$ & $\perb$ & days \\
~~~$e_b$ & $0.22\pm0.13$ & $\eccb$ & \nodata \\
~~~$K_b$ & $1.59 \pm 0.23$ & $\kb$ & m s$^{-1}$ \\[0.8ex]
\hline\\[-2.5ex]
\multicolumn{2}{l}{\cite{Stock:2020}} & & \\
~~~$M_b \sin{i}$ & $2.69\pm0.25$ & $\msinib$ & \mearth \\
~~~$a_b$ & $0.07890^{+0.00068}_{-0.00077}$ & $\ab$ & au \\
~~~$P_b$ & $12.946\pm0.005$ & $\perb$ & days \\
~~~$e_b$ & $0.12^{+0.12}_{-0.09}$ & $\eccb$ & \nodata \\
~~~$P_c$ & $2852\pm568$ & $\perc$ & days \\[0.8ex]
\hline\\[-2.5ex]
\multicolumn{2}{l}{\cite{Rosenthal:2021}} & & \\
~~~$M_b \sin{i}$ & $2.803^{+0.292}_{-0.311}$ & $\msinib$ & \mearth \\
~~~$a_b$ & $0.07892^{+0.00055}_{-0.00098}$ & $\ab$ & au \\
~~~$e_b$ & $0.095^{+0.099}_{-0.066}$ & $\eccb$ & \nodata \\
~~~$M_c \sin{i}$ & $18.0^{+2.9}_{-2.6}$ & $\msinic$ & \mearth \\
~~~$a_c$ & $3.1^{+0.13}_{-0.11}$ & $\ac$ & au \\
~~~$P_c$ & $3190^{+200}_{-170}$ & $\perc$ & days \\
~~~$e_c$ & $0.14^{+0.160}_{-0.095}$ & $\eccc$ & \nodata \\
~~~$P_{cand}$ & $214.59^{+0.99}_{-0.64}$ & $\perd$ & days \\
~~~$K_{cand}$ & $1.45^{+0.42}_{-0.27}$ & $\kd$ & m s$^{-1}$ \\[0.8ex]
\enddata
\end{deluxetable}